\documentclass[%
 reprint,
 amsmath,amssymb,
 aps,
prb,
]{revtex4-1}
\usepackage[utf8]{inputenc}
\usepackage{color}
\usepackage[usenames,dvipsnames,svgnames,table]{xcolor} 
\usepackage{amssymb}
\usepackage{amsmath}
\usepackage{mathtools}
\usepackage{makeidx}
\usepackage{amsfonts}
\usepackage{bm}
\usepackage{rotating}
\usepackage{multirow}
\usepackage{placeins} 
\usepackage{graphicx}
\usepackage{dcolumn}
\usepackage{bm}
\usepackage{hyperref}
\usepackage{xfrac} 
\usepackage[version=4]{mhchem} 

\DeclareMathOperator{\Tr}{Tr}
\providecommand{\e}[1]{\ensuremath{\times 10^{#1}}}
\providecommand{\sllhe}[0]{\ce{^{6}_{\Lambda\Lambda}He}}
\usepackage{feynmp-auto}
\ifpdf
\fi
\makeatletter
\def\endfmffile{%
	\fmfcmd{\p@rcent\space the end.^^J%
			end.^^J%
			endinput;}%
	\if@fmfio
    \immediate\closeout\@outfmf
  \fi
  \IfFileExists{\thefmffile.mp}{\immediate\write18{mpost \thefmffile}}{}
  \let\thefmffile\relax
}
\makeatother

\DeclareMathSymbol{\comma}{\mathpunct}{letters}{"3B}
\makeatletter 
\renewcommand{\maketag@@@}[1]{\hbox{\m@th\normalsize\normalfont#1}}%
\makeatother
\begin{document}

\title{Effects of \texorpdfstring{$\Lambda \Lambda - \Xi {\rm N}$}{LL-XN} mixing in the decay of \texorpdfstring{$S=-2$}{S=-2} hypernuclei}


\author{Jordi Maneu}
\author{Assumpta Parreño}
\author{\`Angels Ramos}
\affiliation{Departament de F\'{\i}sica Qu\`antica i Astrof\'{\i}sica and Institut de Ci\`encies del Cosmos,\\
Universitat de Barcelona, Mart\'{\i} i Franqu\`es, 1, E-08028 Barcelona, Spain}


\date{\today}

\begin{abstract}
We study the non-mesonic weak decay of the doubly-strange hypernucleus \sllhe{} within a model which considers the exchange of pseudoscalar and vector mesons. Special attention is paid into quantifying the strong interaction effects, focussing on the interaction among the two $\Lambda$ hyperons which induces novel weak transitions, whereby $\Lambda \Lambda$, $\Xi N $ and $\Sigma \Sigma$ states decay into a hyperon-nucleon pair. 
The initial strangeness $-2$ wave function is obtained from the solution of a G-matrix equation with the input of realistic strong baryon-baryon potentials, while the final hyperon-nucleon wave functions are derived analogously from a microscopic T-matrix calculation. 
The new $\Lambda\Lambda \to YN$  decay rate studied in this work, $\Gamma_{\Lambda n} + \Gamma_{\Sigma^0 n} + \Gamma_{\Sigma^- p}$, represents  3-4\% of the total one-baryon induced non-mesonic decay and is remarkably affected by strong interaction effects. In particular, the relative importance of the partial decay rates, encoded in the ratio $\Gamma_{\Lambda n} / (\Gamma_{\Sigma^0 n} + \Gamma_{\Sigma^- p})$, gets inverted when the mixing to $\Xi N$ states is incorporated in the initial correlated $\Lambda\Lambda$ wave function. This sensitivity can be used experimentally to learn about the strong interaction in the strangeness $-2$ sector.
\end{abstract}

\pacs{}

\maketitle

\section{Introduction}
Since the discovery in 1952 of the first strange fragment in emulsion chamber experiments, many efforts have been put in extending our knowledge of the nuclear chart towards the SU(3) sector. Worldwide, the study of the interactions among nucleons and hyperons has been a priority in the research plan of many experimental facilities. After more than sixty years of $\Lambda$-hypernuclear studies, remarkable efforts have been devoted to characterise doubly strange systems, with the production of $\Lambda\Lambda$-hypernuclei and more recently, with proposals to study $\Xi$-hypernuclear spectroscopy.
The most effective way of producing doubly strange hypernuclei is through the $(K^{-},K^{+})$ reaction,
which transfers two strangeness and charge units to the target nucleus. Employing high intensity $K^-$ beams of 1.8 MeV/c and high resolution spectrometers, the E05 experiment\cite{Kanatsuki:2015tpa} at J-PARC  plans on producing $^{12}_\Xi$Be hypernuclei, with the goal of studying their spectroscopy, as well as obtaining information on the $\Xi$ potential depth and the $\Xi N\to\Lambda\Lambda$ conversion width. The E07 experiment \cite{NAKAZAWA201569} at J-PARC  aims at producing double-$\Lambda$ hypernuclei in emulsion, following the capture of $\Xi^-$ hyperons at rest, with ten times more statistics than the KEK-E373 experiment\cite{Ahn:2013poa}  that led to the observation of  the \sllhe{} hypernucleus (Nagara event) which established the mild attractive character of the $\Lambda\Lambda$ interaction. Note that a recent re-analysis\cite{doi:10.1093/ptep/ptv008,Sun:2016tuf} of the KEK-E373 experiment, also provided direct evidence for the existence of a bound $\Xi^-$ hypernucleus $^{15}_\Xi$C (Kiso event), produced in the reaction $\Xi^-+^{14}\mathrm{N}\to^{15}_\Xi \mathrm{C}\to^{10}_\Lambda \mathrm{Be}+^5_\Lambda \mathrm{He}$.

Once produced, these strange baryons, unstable with respect to the weak interaction, decay through reactions that do not conserve neither parity, strangeness nor isospin. Being $\Lambda$ the lightest hyperon, its weak decay modes in free space, $\Lambda\rightarrow N\pi$, have been measured with good precision. These mesonic decay channels show, for instance, that transitions which involve a change in isospin of 3/2 are suppressed with respect to those involving a 1/2 variation, a phenomenon known as the $\Delta I=$1/2 rule. Although its origin is not well understood yet, its validity is assumed in most of the theoretical studies of 
weak processes involving hadrons.
In a hypernucleus, the $\Lambda$ is embedded in the nuclear medium, and its mesonic decay mode becomes Pauli blocked as the number of baryons increases, since the emerging nucleon tries to access momentum states that are essentially occupied by the surrounding nucleons. Under such circumstances, new decay mechanisms involving more baryons and with no mesons in the final state start playing an important role. The dominant decay mode for single-$\Lambda$ hypernuclei with $A>5$ is through the two-body process $\Lambda N\to NN$, whereas for double-$\Lambda$ hypernuclei additional hyperon-induced mechanisms, $\Lambda\Lambda\to\Lambda n$, $\Lambda\Lambda\to\Sigma^{-}p$ and $\Lambda\Lambda\to\Sigma^{0}n$ (globally referred to as $\Lambda\Lambda\to YN$ henceforth), become possible. 

Therefore, hypernuclei constitute not only a convenient framework to obtain information on the strong baryon-baryon interaction, but also, through their decay, and using the change of strangeness as signature, a suitable scenario to access both components of the four-fermion weak interaction, the parity-conserving (PC) and parity-violating (PV) amplitudes. 
The large amount of experimental and theoretical work in the strangeness $-1$ sector (see the reviews \cite{Oset:1998xz,Alberico:2001jb,Botta:2012xi}) has led to a consistent interpretation of the experimental data\cite{Ajimura:1997pr,Hashimoto:2002ts,Kim:2003ae,Okada:2004tr,Agnello:2011zza,Agnello:2014sni,Botta:2015lms}, which includes not only lifetimes but also partial decay widths and asymmetries in the angular distribution of the emitted particles.
Crucial elements for this success have been the inclusion of a scalar-isoscalar component \cite{Sasaki:2004xr,Chumillas:2007gi,Itonaga:2008zza,PerezObiol:2011zz} ,  in the weak decay mechanism, or the consideration of medium effects \cite{Ramos:1996ik,Alberico:2004ay,Garbarino:2002wn,Garbarino:2003yq,Bauer:2011gn,Bauer:2017cjk} in the decay observables through the incorporation of multiabsorption processes and 
final state interactions in the data analysis, among others. 
Little is known, comparatively, about the double strangeness systems, due to the small yields for binding two $\Lambda$ particles after $\Xi^-$ capture and the ambiguities in interpreting events owing to the formation of particle-stable excited species. However, this situation might change with the planned experiments at J-PARC which will have much higher statistics.  

Although the new hyperon-induced decay mechanisms provide novel information to constrain the theoretical models, only a limited number of theoretical groups have calculated decay rates for double-$\Lambda$ hypernuclei \cite{Parreno:2001xv,Sasaki:2003nm,Bauer:2015ega}. With the present work we quantify, for the first time, the contribution to the decay of the \sllhe{} hypernucleus of the 
new decay modes, which emerge when the strong interaction is carefully taken into account. Specifically, the weak decay process may occur from a $\Lambda\Lambda$, $\Xi N$ or $\Sigma\Sigma$ state, which can be excited via the strong interaction from the initial $\Lambda\Lambda$ pair. The strong interaction also determines the final $YN$ wave function component, which may have transitioned from either a  $\Lambda N$ or a $\Sigma N$ intermediate state.


For the weak transition we employ a meson-exchange model, built upon the exchange of mesons belonging to the ground state of pseudoscalar and vector octets, thoroughly employed for single-$\Lambda$  hypernuclei, and extended to the decay of double-$\Lambda$ hypernuclei in Ref.~\cite{Parreno:2001xv}. The tree-level values for the baryon-baryon-meson coupling constants are derived using SU(3) symmetry for pseudoscalar mesons and the Hidden Local Symmetry for vector mesons. In the computation of the decay rate, the effects of the strong interaction on the initial state are introduced through the solution of a G-matrix equation, with the input of realistic baryon-baryon potentials\cite{Stoks:1999bz}, while the final hyperon-nucleon wave functions are obtained in an analogous way, by solving the corresponding T-matrix equation.
The essential development with respect to previous calculations \cite{Parreno:2001xv} is the consideration of the weak decay processes from those intermediate states than can be coupled to the initial $\Lambda\Lambda$. This requires the use of  G-matrix wave functions for the coupled transitions  $\Lambda\Lambda$-$\Lambda\Lambda$ and $\Lambda\Lambda$-$\Xi$N. We will see that the $\Lambda\Lambda$-$\Sigma\Sigma$ component is very small and will be neglected in our calculations. Furthermore, the transition potential for the weak $\Xi N-YN$ decay, where $Y$ can be either the $\Lambda$ or $\Sigma$ baryon,  requires the derivation of novel decay constants. These two new ingredients have allowed us to obtain an update on the decay rate for the $(\Lambda\Lambda-\Lambda\Lambda)\to (YN-Y^\prime N^\prime)$ channel as well as new results for the $(\Lambda\Lambda-\Xi N)\to (YN-Y^\prime N^\prime)$ channel, where $Y'$ follows the same criteria as $Y$. 

The manuscript is organized as follows. In Sect.~\ref{rate} we present the formalism used to decompose the hypernuclear decay amplitude for \sllhe{} in terms of two-body $\Lambda\Lambda \to Y N$ amplitudes. The details on how the final $YN$ and initial $\Lambda\Lambda$ wave functions are built are given in Sect.~\ref{sec:gtmatrix}, where results for the different components of the initial state wave function are explicitly shown. The derivation of the isospin structure of the non-relativistic one-meson exchange potential employed in the present work is described in Sect.~\ref{sec:ome}, while the description of the Lagrangians and methodology for obtaining the coupling constants at the strong and weak vertices is relegated to the appendix. Our results are shown in Sect.~\ref{sec:results} and some concluding remarks are given in Sect.~\ref{sec:conclusions}.

\section{Hypernuclear decay rate}
\label{rate}
The nonmesonic decay rate of a hypernucleus is given by
\begin{align}
\Gamma_{nm}=\int\frac{d^3k_1}{(2\pi)^3}\int\frac{d^3k_2}{(2\pi)^3}\sum_{\mathclap{\substack{M_I\left\lbrace R\right\rbrace\\\left\lbrace 1\right\rbrace\left\lbrace 2\right\rbrace}}}&(2\pi)\delta(M_H-E_R-E_1-E_2) \nonumber \\
&\times \frac{1}{2J+1}\vert\mathcal{M}_{fi}\vert^2,
\end{align}
where $M_H$, $E_R$, $E_1$ and $E_2$ correspond to the mass of the hypernucleus, the energy of the $(A-2)$-particle system, and the total asymptotic energies of the emitted baryons, respectively. The integration variables $\vec{k}_1$ and $\vec{k}_2$ stand for the momenta of the two particles in the final state. The momentum conserving delta function has been used to integrate over the momentum of the residual nucleus. The sum, together with the factor $1/(2J+1)$, indicates an average over the initial hypernucleus spin projections, $M_I$, and a sum over all quantum numbers of the residual $(A-2)$-particle system, $\left\lbrace R\right\rbrace$, as well as the spin and isospin projection of the emitted final particles, $\left\lbrace 1\right\rbrace$ and $\left\lbrace 2\right\rbrace$ (henceforth referred to as $\overline{\sum}$). $\mathcal{M}_{fi}$ stands for the transition amplitude from an initial hypernuclear state  (\sllhe{}  in the present study) to a final state composed of a residual nuclear core plus two outgoing baryons. 
When a transformation to the total momentum, $\vec{P}=\vec{k}_1+\vec{k}_2$, and relative momentum, $\vec{k}=(\vec{k}_1-\vec{k}_2)/2$, of the two outgoing particles is performed the expression for $\Gamma_{nm}$ becomes
\begin{align}
\Gamma_{nm}&=\int\frac{d^3P}{(2\pi)^3}\int\frac{d^3k}{(2\pi)^3}\nonumber\\%
&\times\overline{\sum}(2\pi)\delta(M_H-E_R-E_1-E_2)\vert\mathcal{M}_{fi}\vert^2.
\end{align}

We will write the hypernuclear transition amplitude, $\mathcal{M}_{fi}$, in terms of elementary two-body transitions, $B_1B_2\rightarrow B'_{1}B'_{2}$, by using a shell-model framework where $\Lambda$ hyperons and nucleons are described, in a first approximation, by harmonic oscillator single-particle orbitals. The oscillator parameter of the $\Lambda$ particle ($b_\Lambda= 1.6$ fm)  has been chosen to simulate the uncorrelated $\Lambda\Lambda$  probability
of Ref.~\cite{Caro:1998kb} reproducing hypernuclear binding energies, while that of the nucleon ($b_N= 1.4$ fm) is fixed by the $^4$He charge form factor, which determines the size of the nuclear core. In addition, a weak coupling scheme is assumed, by virtue of which the $\Lambda$ hyperons couple only to the ground state of the nuclear core. Therefore, in the case of the \sllhe{} hypernucleus studied here, with spin, isospin quantum numbers $J_I=M_I=0$, $T_I=M_{T_I}=0$, the state will be given by
\begin{align}
\vert \ce{^{6}_{\Lambda\Lambda}He} \rangle=\vert\Lambda\Lambda\rangle^{J=M=0}_{T=M_T=0}\otimes\vert^{4}{\rm He}\rangle^{J_c=M_c=0}_{T_c=M_{T_c}=0},
\end{align}
where antisymmetry forces the two $\Lambda$ hyperons to be in a $^1S_0$ state, since they are assumed to be in the lowest s-shell ($1s_{1/2}$) before the decay occurs. This is so because, in general, a $\Lambda$ in an excited orbital will rapidly decay into the ground state through electromagnetic or strong de-excitation processes, which are orders of magnitude faster than those mediated by the weak interaction.

The evaluation of the $\Lambda N\to  NN$ transition rate requires to decompose the nonstrange nuclear core in terms of one nucleon
coupled to a three-particle system, while decoupling one of the two $\Lambda$ particles, so that the
initial $\Lambda N$  pair can convert into a
final $NN$ pair, conveniently antisymmetrized with the residual nuclear core. The details and final expression for the hypernuclear decay amplitude in terms of two-body $\Lambda N \to NN$ ones can be found in Ref.~\cite{Parreno:2001xv}. Here, we focus on the $\Lambda \Lambda \to YN$ decay mode, which is the one we improve with respect to earlier calculations. In this case, the $\Lambda$ hyperon does not need to be decoupled from the cluster, neither a nucleon from the core. The residual 4-particle system, which coincides with the $^4$He nucleus, contains no strangeness, while the final two-particle state contains one hyperon than can be either a $\Lambda$ ($\vert Yt_Y\rangle=\vert 00\rangle$), a $\Sigma^-$ ($\vert Y t_{Y}\rangle=\vert 1-1\rangle$) or a $\Sigma^0$ ($\vert Y t_{Y}\rangle=\vert 10\rangle$). The hypernuclear amplitude associated to the $\Lambda\Lambda\rightarrow YN$ transition is then given by
\begin{align}
&\mathcal{M}_{\Lambda\Lambda\rightarrow YN}=\langle \vec{k}_Ns_Nt_N,\vec{k}_Ys_Yt_Y;^4{\rm He}\vert \hat{O}\vert \ce{^{6}_{\Lambda\Lambda}He}\rangle\vert\tfrac{1}{2}-\tfrac{1}{2}\rangle_{\Lambda}\nonumber\\
&=\sum_{S^\prime,M^\prime_S}\langle\tfrac{1}{2}s_N,\tfrac{1}{2}s_Y\vert S^\prime M^\prime_S\rangle\langle\tfrac{1}{2}t_N,Yt_Y\vert T^\prime M^\prime_T\rangle\nonumber\\
&\times\langle\vec{K}\vert\Psi^{CM}_{\Lambda\Lambda}\rangle\langle\vec{k}Y, S^\prime M^\prime_S,T^\prime M^\prime_T\vert\hat{O}\vert\Psi^{rel}_{\Lambda\Lambda} ,S M_{S},T M_{T}\rangle \ ,
\label{eq:transition}
\end{align}
where the initial $\Lambda \Lambda$ wave function has been written as a product of  relative and  center-of -mass wave-functions, $\Psi^{rel}_{\Lambda\Lambda}$ and $\Psi^{CM}_{\Lambda\Lambda}$, respectively, and $\vec{k}$ and $\vec{K}$ are the relative and total momentum of the emitted $YN$ pair.
The amplitude $\langle\vec{k},Y S^\prime M^\prime_S,T^\prime M^\prime_T\vert\hat{O}\vert\Psi^{rel}_{\Lambda\Lambda},S M_{S},T M_{T}\rangle$ represents the two-body transition matrix element.  
The spin quantum numbers of the $\Lambda\Lambda$ pair are $S=M_{S}=0$, due to antisymmetrization,   while its isospin quantum numbers are $T=1/2$, $M_{T}=-1/2$, which contain the coupling to the isospurion field $\vert\tfrac{1}{2}-\tfrac{1}{2}\rangle_{\Lambda}$ introduced to account for the $\Delta I=1/2$ rule in the transition. The isospin quantum numbers of the emitted pair fulfill $T^\prime=T=1/2$ and  $M^\prime_T=M_{T}=-1/2$ by isospin conservation.

These two-body matrix elements are calculated using a two-body interaction potential based on a meson-exchange picture, including initial and final correlated wave functions which are derived through the resolution of the G-matrix formalism for the former and the Lippmann-Schwinger equation for the latter. Further details will be given in the following sections.

In the case of the $\Lambda\Lambda \to YN$ decay, and since the strong interaction allows for the conversion to other baryon-baryon channels, the weak interaction will take place not only from the initial $\Lambda\Lambda$ pair, but also from pairs containing other members of the baryon octet, $\Xi^{-}p$, $\Xi^0 n$, $\Sigma^0 \Sigma^0$ and $\Sigma^{+}\Sigma^{-}$, which will in turn decay weakly into either $\Lambda n$, $\Sigma^{0}n$ or $\Sigma^- p$ states. Moreover, the strong interaction acting between the final baryons will produce additional $Y^\prime N - YN$ transitions, with $Y^\prime,Y$ being $\Lambda$ or $\Sigma$ hyperons . 

\begin{figure}[!ht]
	\centering
	\includegraphics[width=0.25\textwidth]{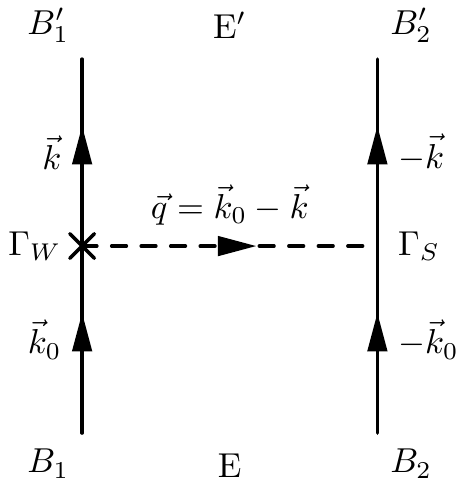}%
\caption{Diagrammatric representation of a  $B_1B_2\to B_1'B_2'$ transition within a one-meson exchange model. The cross represents an insertion of a weak vertex.}
\label{fig:momenta}
\end{figure}

In a meson exchange picture, every weak $B_1B_2\rightarrow B'_{1}B'_{2}$ transition can be understood in terms of the exchange of mesons between the interacting particles, with masses related to the inverse of the interaction range and with quantum numbers allowed by the symmetries governing the underlying dynamics.  Within this model, the transition involves a product of a strong vertex and a weak one, where the change in strangeness occurs, connected through the meson propagator (see Fig.~\ref{fig:momenta}).

We note that, while the calculation corresponding to the exchange of pseudoscalar mesons other than the pion requires the use of SU(3)$_f$ symmetry to obtain the baryon-baryon-meson vertices, the inclusion of vector mesons requires the generalization of the Hidden Local Symmetry (HLS) formalism\cite{HLS-85} to the strange strong sector, and the implementation of SU(6)$_W$ for the weak vertices, as explained in the appendix. One might also use the less model dependent effective field theory approach to describe the four-fermion weak interaction, which would replace those vector meson exchanges by contact terms \cite{Jun:2001mn,Parreno:2003ny,PerezObiol:2011zz,Perez-Obiol:2013waa}. These contributions are organised
as an expansion of increasing dimension (powers of some {\it small} ratio of physical scales), providing a more systematic and controllable framework to study the weak process. The size of the coefficients in the expansion is constrained by fitting to accurate experimental data. At present, there are no measurements for the required weak transitions in the strangeness $-2$ sector, and one has to rely on model determinations of the decay
mechanism as the meson-exchange approach employed in the present work.

\section{\label{sec:gtmatrix}Strong interactions}

In this section we discuss how to incorporate the strong interaction effects into the calculation of the weak decay amplitude.
We employ the $G$-matrix formalism to obtain the correlated wave function for the initial hyperon-hyperon state, which takes into account the Pauli blocking effects on the nucleon of the intermediate $\Xi N$ pair. For the description of the final $YN$ system, the relevant strong interaction effects in the evaluation of decay rates are those associated with their mutual interaction, which can be addressed by solving the corresponding $T-$matrix equation. The propagation of these two baryons through the residual nuclear medium induces additional final state interactions, which modify energy and angular spectra of the emitted particles.  Using an intranuclear cascade code (Monte-Carlo), one can then compare to experimental data at the level of the detected particle spectra. Since our interest with the present calculation is to estimate decay rates, we will omit the involved analysis of the particle spectra, and focus only on the interaction between the two weakly emitted baryons.
%
Note that, given the angular quantum numbers of the $\Lambda\Lambda$ initial state, $^1S_0$, the possible $^{2S+1}L_J $ channels involved in the process $\Lambda\Lambda \to B_1 B_2 \to B^\prime_1 B^\prime_2 \to YN$ are listed in Table~\ref{tab:transitions}, where conservation of total angular momentum, as well as conservation of parity for the strong interaction transitions, has been taken into account. The symbols in between brackets denote the labels that will represent the different baryon-baryon channels in the two-body states.
\begin{table}
\caption{Possible $^{2S+1}L_J$ channels involved in the weak process $\Lambda\Lambda \to B_1 B_2 \to B^\prime_1 B^\prime_2 \to YN$ contributing to the weak decay of \sllhe{}, where the first and last transitions are mediated by the strong interaction.}
\begin{ruledtabular}
\begin{tabular}{ccccccc}
 $\Lambda\Lambda$ & $\rightarrow$ & $B_1$ $B_2$ & $\rightarrow$  & $B_1^\prime$ $B_2^\prime$ & $\rightarrow$ & $YN$  \\
   ($\Lambda$) & & $(\tilde{Y})$ & &  $(Y^\prime)$ &  & $(Y)$ \\
\hline
\\
$^{2S+1}L_J$ & & $^{2{\tilde S}+1}{\tilde L}_J $ & & $^{2{\tilde S}^\prime+1}{\tilde L}^\prime_J $ & & $^{2S^\prime+1}L^\prime_J $ \\
\hline
\\
$^1S_0$ & $\rightarrow$ & $^1S_0$ & $\rightarrow$  & $^1S_0$  & $\rightarrow$& $^1S_0$ \\
$^1S_0$ & $\rightarrow$ & $^1S_0$ & $\rightarrow$  & $^3P_0$ & $\rightarrow$  & $^3P_0$ 
\end{tabular}
\end{ruledtabular}
\label{tab:transitions}
\end{table}

\subsection{Final State}
The effect of the strong interaction between the outgoing hyperon and nucleon can be accounted for by the $YN$ wave functions. Let the Hamiltonian be  $H=H_0+V$.
If we denote by $\Phi$ the plane wave solution, $\vert\vec{k}Y S^\prime M^\prime_S\rangle$, of the Hamiltonian $H_0$ with energy E, \textit{i.e.} $H_0\Phi=E\Phi$, then the possible solutions for $\Psi$ are given by the Lippmann-Schwinger equation,
\begin{align}
\vert\Psi^{(\pm)}\rangle=\vert\Phi\rangle+\frac{V\vert\Psi^{(\pm)}\rangle}{E-H_0\pm i\epsilon},
\end{align}
where the positive (negative) solution corresponds to a plane wave plus an outgoing (incoming) spherical wave at sufficiently large distances. An alternative formulation of the Lippmann-Schwinger equation written in terms of the transition matrix $T$ yields
\begin{gather}
\vert\Psi^{(+)}\rangle=\vert\Phi\rangle+\frac{T\vert\Phi\rangle}{E-H_0+i\epsilon},\\
\label{eq:Outgoing}\langle\Psi^{(-)}\vert=\langle\Phi\vert+\frac{\langle\Phi\vert T}{E-H_0-i\epsilon},
\end{gather}
where the T operator fulfills
\begin{align}
T=V+V\frac{T}{E-H_0+i\epsilon}.
\end{align}

Projecting into coordinate space and inserting a complete set of states on the r.h.s in Eq.~(\ref{eq:Outgoing}) we find:
\begin{align}
&\langle\Psi^{(-)}_{\vec{k}} YS^\prime M^\prime_S  \vert\vec{r}\,\rangle=\langle\vec{k} Y S^\prime M^\prime_S \vert\vec{r}\,\rangle 
 \nonumber\\%
&+\sum_{\tilde{S}^\prime \tilde{M}^\prime_S} \sum_{Y^\prime} \int d^3 k^\prime \frac{\langle\vec{k} Y S^\prime  M^\prime_S \vert T\vert\vec{k}^\prime Y^\prime \tilde{S}^\prime  \tilde{M}^\prime_S \rangle\langle\vec{k}^\prime Y^\prime \tilde{S}^\prime  \tilde{M}^\prime_S  \vert\vec{r}\,\rangle}{E-H_0-i\epsilon}.
\label{eq:wavef}
\end{align}
We perform a partial wave decomposition in the coupled $(LS)J$ representation of the wave functions $\langle\Psi^{(-)}_{\vec{k}} Y S^\prime M^\prime_S \vert\vec{r}\,\rangle$ and $\langle\vec{k} Y S^\prime M^\prime_S \vert\vec{r}\,\rangle$, the latter being the adjoint of the free plane wave, $\mathrm{e}^{-i\vec{k}\vec{r}}\langle Y S^\prime M^\prime_S \vert$, and obtain:
\begin{align}
\label{eq:WF}
&\Psi^{(-)*}_{\vec{k}\,Y}(\vec{r}\,) \chi^{S^\prime}_{M^\prime_S}= \nonumber \\
&~~~~~4\pi\sum_{JM}\sum_{L^\prime M^\prime_L \tilde{L}^\prime \tilde{S}^\prime} \sum_{Y^\prime} (-i)^{\tilde{L}^\prime} \Psi^{(-)*J}_{Y L^\prime S^\prime,Y^\prime\tilde{L}^\prime \tilde{S}^\prime}(k,r)Y_{L^\prime M^\prime_L}(\hat{k})\nonumber\\%
&~~~~~\times\langle L^\prime M^\prime_L S^\prime M^\prime_S\vert JM\rangle\mathcal{J}^{\dagger JM}_{\tilde{L}^\prime \tilde{S}^\prime}(\hat{r}),
\end{align}
where the generalized spherical harmonic  $\mathcal{J}^{\dagger}$ is defined as 
\begin{align}
\mathcal{J}^{\dagger JM}_{\tilde{L}^\prime \tilde{S}^\prime}(\hat{r})=\sum_{\tilde{M}^\prime_L \tilde{M}^\prime_S}\langle \tilde{L}^\prime \tilde{M}^\prime_{L} \tilde{S}^\prime \tilde{M}^\prime_{S} \vert JM\rangle Y^{*}_{ \tilde{L}^\prime \tilde{M}^\prime_{L}}(\hat{r}).
\end{align}
The partial wave decomposition for the free plane wave may be obtained by replacing $\Psi^{(-)*J}_{Y L^\prime S^\prime,Y^\prime\tilde{L}^\prime \tilde{S}^\prime}(k,r)$ with $j_{L^\prime}(kr)\delta_{Y^\prime Y}\delta_{\tilde{L}^\prime L^\prime}\delta_{\tilde{S}^\prime S^\prime}$ in Eq.~(\ref{eq:WF}), where $j_{L^\prime}(kr)$ is the spherical Bessel wave function. 

For the $T$ matrix elements one can write
\begin{align}
\langle\vec{k} Y S^\prime M^\prime_S \vert T & \vert\vec{k}^\prime Y^\prime \tilde{S}^\prime \tilde{M}^\prime_S \rangle= \nonumber \\
& \sum_{JM}\sum_{L^\prime M^\prime_L}\sum_{\tilde{L}^\prime \tilde{M}^\prime_L}Y_{L^\prime M^\prime_L}(\hat{k})Y^{*}_{\tilde{L}^\prime M^\prime_{\tilde{L}}}(\hat{k}^\prime)\nonumber\\%
&\times\langle L^\prime M^\prime_L S^\prime M^\prime_S \vert JM\rangle\langle \tilde{L}^\prime \tilde{M}^\prime_L \tilde{S}^\prime \tilde{M}^\prime_S \vert JM\rangle\nonumber\\%
&\times\langle k Y (L^\prime S^\prime)JM\vert T\vert k^\prime Y^\prime(\tilde{L}^\prime \tilde{S}^\prime)JM\rangle \ .
\end{align}
Inserting the previous equation, together with the partial wave decomposition of the wave-functions in Eq.~(\ref{eq:wavef}), one obtains the equation that determines the partial wave components of the correlated wave function:
\begin{align}
&\Psi_{Y L^\prime S^\prime,Y^\prime \tilde{L}^\prime \tilde{S}^\prime}^{(-)*J}(k,r)=j_{L^\prime}(kr)\delta_{Y^\prime Y} \delta_{\tilde{L}^\prime L^\prime}\delta_{\tilde{S}^\prime S^\prime}+\nonumber\\%
&\int k^{\prime\,2}dk^\prime\frac{\langle k Y (L^\prime S^\prime)JM\vert T\vert k^\prime Y^\prime (\tilde{L}^\prime \tilde{S}^\prime)JM \rangle j_{\tilde{L}^\prime}(k^\prime r)}{E(k)-E(k^\prime)+i\epsilon} \ ,
\end{align}
where the partial wave $T$-matrix elements fulfil the integral equation:
\begin{align}
& \langle k Y (L^\prime S^\prime)JM \vert T \vert k^\prime Y^\prime (\tilde{L}^\prime \tilde{S}^\prime)JM \rangle \nonumber\\
& =\langle k Y (L^\prime S^\prime)JM\vert V \vert k^\prime Y^\prime (\tilde{L}^\prime \tilde{S}^\prime)JM\rangle\nonumber\\%
&+\sum_{L^{\prime\prime}S^{\prime\prime}Y^{\prime\prime}}\int k^{\prime\prime \, 2}dk^{\prime\prime}\langle k Y (L^\prime S^\prime)JM\vert V\vert k^{\prime\prime} Y^{\prime\prime} (L^{\prime\prime} S^{\prime\prime} )JM\rangle\nonumber\\%
&\times\frac{\langle k^{\prime\prime} Y^{\prime\prime} (L^{\prime\prime} S^{\prime\prime})JM\vert T\vert k^{\prime}Y^\prime(\tilde{L}^\prime \tilde{S}^\prime)JM\rangle}{E(k)-E(k^{\prime\prime})+i\epsilon}\ .
\end{align}

Note that, since  the $\Lambda\Lambda$ pair is in a $^1S_0$ state, conservation of angular momentum and parity prevents a change of the spin and orbital angular momentum quantum numbers between the pre- and post-strong transition states, as seen in Table~\ref{tab:transitions}. Consequently, the above equations could be simplified by applying $L^{\prime\prime} =\tilde{L}^\prime=L^\prime$, $\tilde{M}_{L^\prime}=M_{L^\prime}$, $S^{\prime\prime}= \tilde{S}^\prime=S^\prime$ and $\tilde{M}_{S^\prime}=M_{S^\prime}$. 

\subsection{Initial state}
For the initial state interactions a similar framework to that of the Lippmann-Schwinger equation is applied, but considering the fact that the interacting particles feel the presence of the medium where they are embedded. This is known as the Brueckner-Goldstone theory, which considers the interactions of a pair of particles within the Fermi sea, with the collisions fulfilling the requirements of the Pauli principle. We will solve the problem in infinite nuclear matter and the obtained results will be applied to the finite hypernuclear problem that we are dealing with.

Working within the $\left(\frac{1}{2}\right)^{+}$ baryon octet, the strange $\Lambda\Lambda$, $\Xi N$ and $\Sigma\Sigma$ pairs can couple through the strong interaction to the initial $\Lambda\Lambda$ state. The correlated state, $\vert\Psi\rangle$, is defined through $G\vert\Psi\rangle=V\vert\Phi\rangle$, where $\vert\Phi\rangle$ is the free-particle state, and $G$ is given in terms of the bare baryon-baryon potential, $V$, by:
\begin{align}
G=V+V\frac{Q}{E-H_0+i\epsilon}G.
\end{align}
This is an integral equation, where $Q$ corresponds to the Pauli blocking operator, which restricts the summation to unoccupied states above the Fermi level, and E is the energy of the interacting two-body system. The correlated state can therefore be written as:
\begin{align}
\vert\Psi\rangle=\vert\Phi\rangle+\frac{Q}{E-H_0+i\epsilon}G\vert\Phi\rangle.
\end{align}

Working in the coupled $(LS)J$ representation, we find
\begin{align}
&\Psi^{J}_{\tilde{Y} \tilde{L} \tilde{S},\Lambda LS}(k,r)=j_L(k r)\delta_{\Lambda \tilde{Y}}\delta_{L\tilde{L}}\delta_{S\tilde{S}}\\%
&+\int k^{\prime\,2} dk^{\prime}\frac{\langle k^{\prime} \tilde{Y} (\tilde{L} \tilde{S})JM \vert G\vert k Y\Lambda (LS)JM\rangle\overline{Q}(k^{\prime})j_{\tilde{L}}(k^{\prime}r)}{E(k)-E(k^{\prime})+i\epsilon} \nonumber,
\end{align}
where $\overline{Q}$ stands for the angle-averaged Pauli operator and the partial wave $G$-matrix elements fulfill the integral equation:
\begin{align}
&\langle k^{\prime}  \tilde{Y}  (\tilde{L} \tilde{S})JM\vert G\vert k \Lambda (LS)JM\rangle \nonumber \\
&=\langle k^{\prime} \tilde{Y}(\tilde{L} \tilde{S})JM\vert V\vert k \Lambda (LS)JM\rangle\nonumber\\%
&+\sum_{L^{\prime\prime} S^{\prime\prime} Y^{\prime\prime}}\int k^{\prime\prime\, 2} dk^{\prime\prime}\langle k^{\prime}  \tilde{Y} (\tilde{L} \tilde{S})JM\vert V\vert k^{\prime\prime}Y^{\prime\prime}(L^{\prime\prime}S^{\prime\prime})JM\rangle\nonumber\\%
&\times\frac{\overline{Q}(k^{\prime\prime})\langle k^{\prime\prime} Y^{\prime\prime} (L^{\prime\prime}S^{\prime\prime})JM\vert G\vert k \Lambda (LS)JM\rangle}{E(k)-E(k^{\prime\prime})+i\epsilon}.
\end{align}
As before, conservation of angular momentum and parity, together with the fact that the initial $\Lambda\Lambda$ is in a $^1S_0$ state, simplifies the above equations considerably, as the only permitted transition is $^1S_0 \to ^1S_0$.

In order to obtain the wave functions in a finite hypernucleus a correlation function is defined:
\begin{align}
f^{J}_{\Lambda LS}(r)\equiv\frac{\Psi^{J}_{\Lambda LS,\Lambda LS}(k^{*},r)}{j_{L}(k^{*}r)}.
\label{eq:corr}
\end{align}
This function ascribes the correlated wave function in nuclear matter, $\Psi^{J}_{\Lambda LS,\Lambda LS}(k,r)$, with the non-interacting one for a relative momentum $k^{*}$, taken to be 100 MeV, which is representative of the average momentum of the $\Lambda\Lambda$ pair in \sllhe{}. The same correlation effects are assumed for finite nuclei, thusly defining the diagonal terms of the relative motion wave function in a finite nucleus as:
\begin{align}
\Omega^{J}_{\Lambda LS}(r)\equiv f^{J}_{\Lambda LS}(r)\Phi_{NL}\left(\frac{r}{\sqrt{2}b}\right),
\label{eq:diag}
\end{align}
where $\Phi_{NL}(r/\sqrt{2}b)$ is the relative harmonic oscillator wave function of the two $\Lambda$ particles.

For the non-diagonal $\Lambda LS\to \tilde{Y}\tilde{L}\tilde{S}$ components, we rescale the nuclear matter wave-function $\Psi^{J}_{\tilde{Y} \tilde{L} \tilde{S},\Lambda LS}$ by the same normalisation factor affecting the diagonal components, namely:
\begin{align}
\Omega_{\tilde{Y} \tilde{L} \tilde{S},\Lambda LS}^{J}(r)\equiv\frac{\Phi_{NL}(r=0)}{j_{L}(k^{*}r=0)}\Psi_{\tilde{Y} \tilde{L} \tilde{S},\Lambda LS}^{J}(k^*,r)  ,
\end{align}
as can be inferred upon inspecting Eqs.~(\ref{eq:corr}) and (\ref{eq:diag}).

\begin{figure}[!ht]
	\centering
	\includegraphics[width=0.48\textwidth]{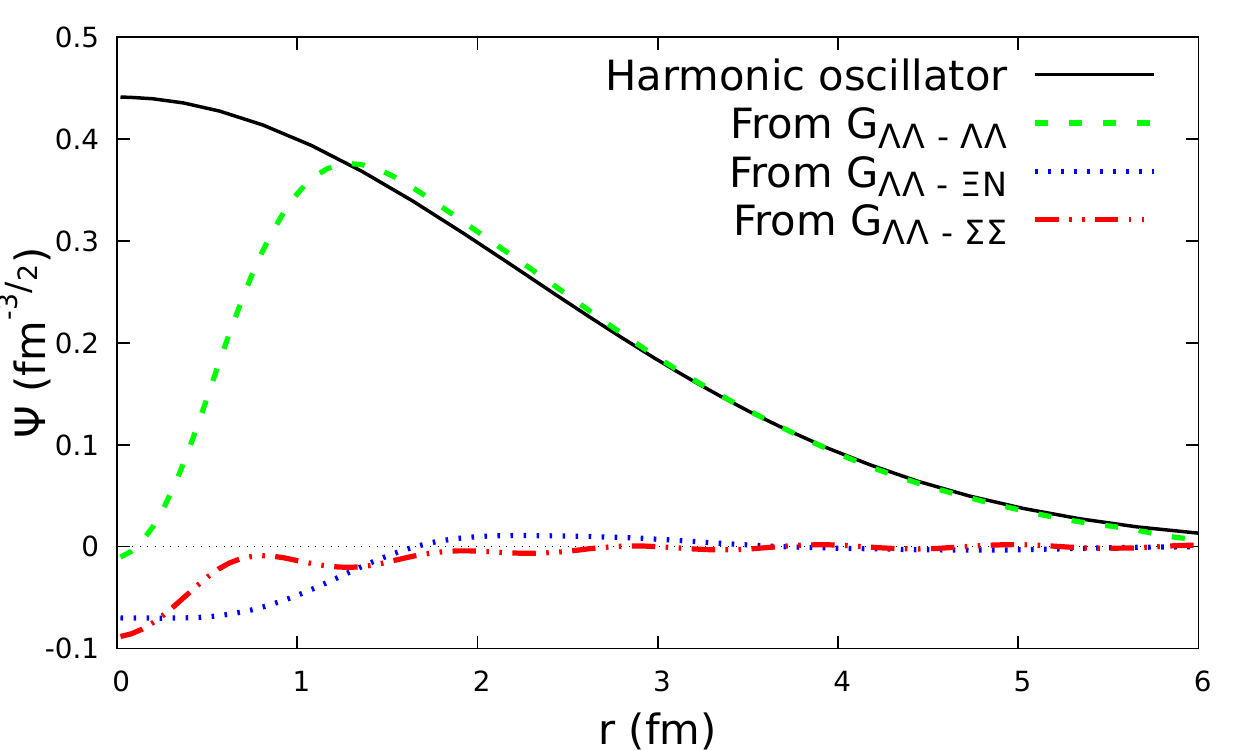}%
	\caption{\label{fig:WF}Different components of the relative $\Lambda \Lambda$ wave function corresponding to the $^1S_0$ channel, and for a value of the relative momentum of 100 MeV.}
\end{figure}

In Fig.~\ref{fig:WF} we represent the initial $\Lambda \Lambda$ wave function for $N=1$ and $L=0$ as a function of the relative distance between the two $\Lambda$ particles. The   
black solid line displays the uncorrelated  harmonic oscillator wave function, while the green dashed line displays the correlated wave function for the dominant diagonal $\Lambda\Lambda-\Lambda\Lambda$ component. The red dot-dot-dashed and the blue dotted lines represent, respectively, the $\Lambda\Lambda-\Sigma\Sigma$ and the $\Lambda\Lambda-\Xi N$ components. It is clear that the dominant contribution to the $\Lambda \Lambda \to YN$ decay mode of \sllhe{} will come from the diagonal $\Lambda\Lambda -\Lambda\Lambda$ component of the wave function, which at large distances behaves as an uncorrelated harmonic oscillator while at short distances its strength gets reduced due to the short-distance repulsive behavior of the strangeness $S=-2$ baryon-baryon NSC97f interaction employed \cite{Stoks:1999bz}. With regard to the non-diagonal components of the wave function one can see that, despite having a comparable size at the origin, the strength of the $\Lambda\Lambda - \Sigma\Sigma$ term is essentially located at distances under 0.5 fm, which will be strongly suppressed by the $r^2$ factor in the integrand of the two-body matrix element. On the other hand, the $\Lambda\Lambda - \Xi N$ component is still sizable around 1 fm and it is expected to contribute non-negligibly to the $\Lambda \Lambda \to YN$ decay mode of \sllhe{}. We will see that, even if non-negligible, this non-diagonal component gives only a ten percent correction to the diagonal contribution, a finding that justifies that in this work we disregard the contributions to the decay coming from the
$\Lambda\Lambda - \Sigma\Sigma$ component of the initial $\Lambda\Lambda$ wave-function.

\section{One-Meson-Exchange Potential}
\label{sec:ome}
The evaluation of the two-body transition matrix elements of Eq.~(\ref{eq:transition}) requires the knowledge of the operator that triggers the weak $\Delta S=-1$ transition from an initial baryon pair to a final one.  
In the meson exchange description employed here these transitions are assumed to proceed via exchange of virtual mesons belonging to the pseudoscalar and vector meson octets. The corresponding transition potential is obtained from the amplitude displayed in Fig.~\ref{fig:momenta}, which is written as
\begin{align}
\label{Eq:M}
\mathcal{M}=\int d^4xd^4y\overline{\Psi}_{1'}(x)\Gamma_{1}\Psi_{1}(x)\Delta_\phi(x-y)\overline{\Psi}_{2'}(y)\Gamma_{2}\Psi_{2}(y),
\end{align}
with $\Gamma_i$ being the Dirac operators characteristic of the baryon-baryon-meson vertices and $\Delta_\phi(x-y)$ the meson ($\phi$) propagator:
\begin{align}
\Delta_\phi(x-y)=\int\frac{d^4q}{(2\pi)^4}\frac{\mathrm{e}^{iq(x-y)}}{(q^0)^2-\vec{q}\,^2-m_\phi^2}.
\end{align}

Combining the above two expressions, performing a change to center-of-mass (c.m.) and relative variables and integrating over the c.m., time and energy variables, one obtains the amplitude in terms of the vertices that come from the matrix elements between fields, $\overline{\Psi}(x)\Gamma\Psi(x)$. In Table \ref{tab:vert} we show the strong and weak $\Gamma$ operators
for pseudoscalar (PS) and vector (V) mesons. The constants $A$, $B$, $\alpha$, $\beta$ and $\epsilon$ correspond to the weak couplings, while $g$ ($g^{\rm V},g^{\rm T}$) represents the strong (vector, tensor) one.
\begin{table}
\caption{\label{T:vert}Pseudoscalar (PS) and vector (V) vertices entering 
Eq.~(\ref{Eq:M}) (in units
of $G_F {m_\pi}^2=2.21 \e{-7}$).}
\begin{ruledtabular}
\begin{tabular}{lcc}
 & PS & V \\
\hline\\
Strong &  $ {\rm i} g  \gamma_5  $ &
$  \left[ g^{\rm V} \gamma^\mu + {\rm i}
\displaystyle\frac{ g^{\rm T}}{2M}\sigma^{\mu \nu} q_\nu \right]  $ \\
 & & \\
\hline\\
 & & \\
Weak & ${\rm i}  (A + B \gamma_5)  $ &
$  \left[\alpha \gamma^\mu - \beta {\rm i}
\displaystyle\frac{\sigma^{\mu \nu} q_\nu}
{2 \overline{M}}+\varepsilon\gamma^\mu \gamma_5 \right] $ \\
 & & \\
\end{tabular}
\end{ruledtabular}
\label{tab:vert}
\end{table}

We take the non-relativistic reduction of this transition amplitude and the static $q^0=0$ limit, allowing us to identify $\mathcal{M}(\vec{q}\,)$ with $V(\vec{q}\,)$, which is the Fourier transform of the transition potential in coordinate space. As we detail below, the general structure of this potential for pseudoscalar meson exchanges reads:
\begin{align}
V^{\phi}(\vec{q}\,)=\sum_{k}\left(A^{Y,\phi}_{k}+\frac{B^{Y,\phi}_{k}}{2\overline{M}}\vec{\sigma}_{1}\vec{q}\,\right)\frac{\vec{\sigma}_{2}\vec{q}}{\vec{q}^{\,2}+m_{\phi}^{2}}\hat{O}^{Y}_{k},
\label{eq:pseudo}
\end{align}
where $\overline{M}$ is the average mass of the baryons involved in the weak vertex, $m_\phi$ is the mass of the exchanged meson and the index $k$ in the sum runs over the different isospin structures associated to each type of meson. Similarly, the potential for vector meson exchanges reads:
\begin{align}
V^{v}(\vec{q}\,)=\sum_{k}\Bigg({\rm i}\frac{A^{Y,v}_{k}}{2M}\left(\vec{\sigma}_{1}\times\vec{\sigma}_{2}\right)\vec{q}+B^{Y,v}_{k}\nonumber\\%
+\frac{C^{Y,v}_{k}}{4M\overline{M}}\left(\vec{\sigma}_{1}\times\vec{q}\,\right)\left(\vec{\sigma}_{2}\times\vec{q}\,\right)\Bigg)\frac{1}{\vec{q}^{\,2}+m_{v}^{2}}\hat{O}^{Y}_{k}.
\label{eq:vector}
\end{align}
The explicit expressions for the $A$, $B$ and $C$ constants, in terms of strong and weak coupling constants, will be given at the end of this section. 

In order to build the isospin operators $\hat{O}^{Y}_{k}$ one needs to know the isospin nature of the meson being exchanged (isoscalar for $\eta$ and $\omega$, isodoublet for $K$ and $K^*$ , and isovector for $\pi$ and $\rho$) and the specific baryons involved in the two-body weak transition. Note that the  $\Delta I=1/2 $ rule is implemented through the insertion of an isospurion $\mid \frac{1}{2}\, -\frac{1}{2}\, \rangle$ in the initial state. We focus on developing the isospin structure for the transitions $\Xi N\to Y N $, with $Y=\Lambda,\Sigma$, which are the new contributions considered in the present work. 
Attending only to the isospin quantum numbers, the general structure of the $\Xi N\to Y N $ matrix element is:
\begin{align}
&g^{Y,\phi}_1 \langle Y\,t_{Y},\tfrac{1}{2}\,t_{Nf}\vert \hat{O}^{Y}_1 \vert 0\,t_{\Xi}-\tfrac{1}{2}, \, \tfrac{1}{2}\,t_{Ni}\rangle\nonumber\\&\times\langle 0\,t_{\Xi}-\tfrac{1}{2}\vert\tfrac{1}{2}\,t_{\Xi},\tfrac{1}{2}\,-\tfrac{1}{2}\rangle\nonumber\\%
+&g^{Y,\phi}_2 \langle Y\,t_{Y},\tfrac{1}{2}\,t_{Nf}\vert\hat{O}^{Y}_2 \vert 1\,t_{\Xi}-\tfrac{1}{2}, \, \tfrac{1}{2}\,t_{Ni}\rangle\nonumber\\&\times\langle 1\,t_{\Xi}-\tfrac{1}{2}\vert\tfrac{1}{2}\,t_{\Xi},\tfrac{1}{2}\,-\tfrac{1}{2}\rangle\nonumber\\%
+&g^{Y,\phi}_3 \langle Y\,t_{Y},\tfrac{1}{2}\,t_{Nf}\vert\hat{O}^{Y}_3 \vert 1\,t_{\Xi}-\tfrac{1}{2}, \, \tfrac{1}{2}\,t_{Ni}\rangle\nonumber\\&\times\langle 1\,t_{\Xi}-\tfrac{1}{2}\vert\tfrac{1}{2}\,t_{\Xi},\tfrac{1}{2}\,-\tfrac{1}{2}\rangle.
\label{eq:isomat}
\end{align}
where the isospurion has been coupled to the isospin $\frac{1}{2}$ of the $\Xi$ giving states with isospin $I=0,1$, which in turn couple to the initial nucleon isospin ($t_{Ni}=\frac{1}{2}$ for a $p$ and $-\frac{1}{2}$ for a $n$) to give the final $YN$ state.

We first examine those cases where the final state is of the $\Lambda N$ type, i.e. $\vert Y t_{Y}\rangle=\vert 0\,0\rangle$. The possible isospin operators can be argued to be
\begin{gather}
\widehat{O}_1^\Lambda\equiv\mathbb{I}_1\otimes\mathbb{I}_2,\\
\widehat{O}_2^\Lambda\equiv\vec{\mathbb{T}}_{01}\otimes\vec{\tau}_2,\\
\widehat{O}_3^\Lambda\equiv  0
\end{gather}
where $\vec{\tau}$ stands for the Pauli matrices and $\vec{\mathbb{T}}_{01}$ is an operator that allows the transition from a $I=1$ state to a $I=0$ one. Their spherical coordinates have the following matrix elements
\begin{align}
\langle 00\vert\mathbb{T}^k_{01}\vert 1m\rangle=(-1)^k\langle 0\,0,1\,-k\vert 1\,m\rangle=(-1)^k\delta_{m,-k}.
\label{eq:tiso01}
\end{align}
We note that, in the case of a $\Lambda N$ final state, we have set the $\widehat{O}_3^\Lambda$ operator to zero to account for  the fact that there is only one possible scalar operator ($\widehat{O}_2^\Lambda$) connecting the initial $\vert 1\,t_{\Xi}-\tfrac{1}{2}, \, \tfrac{1}{2}\,t_{Ni}\rangle$ pair with the final $\langle 0\,0,\tfrac{1}{2}\,t_{Nf}\vert$ one.
In the case of a $\Sigma N$ final state, we have $\vert Y t_{Y}\rangle=\vert 1 t_{Y}\rangle$ and the corresponding appropriate set of operators is:
\begin{align}
\widehat{O}_1^\Sigma\equiv\vec{\mathbb{T}}_{10}\otimes\vec{\tau}_2,\\
\widehat{O}_2^\Sigma\equiv\mathbb{I}_1\otimes\mathbb{I}_2,\\
\widehat{O}_3^\Sigma\equiv\vec{\mathbb{T}}_{11}\otimes\vec{\tau}_2 ,
\end{align}
where $\vec{\mathbb{T}}_{10}$ mediates transitions from isospin 0 to isospin 1, with matrix elements 
\begin{align}
\langle 1 m \vert\mathbb{T}^k_{10}\vert 00 \rangle=\delta_{m,k} \ ,
\end{align}
as can be inferred taking the adjoint in Eq.~ (\ref{eq:tiso01}).
Likewise $\vec{\mathbb{T}}_{11}$ induces transitions from an initial $I=1$ state to a final $I=1$ one. Its matrix elements are given by:
\begin{align}
\langle 1 m' \vert\mathbb{T}^k_{11}\vert 1 m \rangle=\sqrt{2} \langle 1 m' \vert 1 m 1 k \rangle \ .
\end{align}

In the following, we will write the isospin coefficients $g^{Y,\phi}_i$ ($i=1,2,3$) in terms of the weak and strong coupling constants characteristic of the meson exchanged in either the $\Xi N \to \Lambda N$ or $\Xi N \to \Sigma N$ process. The explicit expressions for the weak and strong coupling constants can be found in the appendix.

In the case of the isoscalar mesons ( $\eta$ or $\omega$ ) only the $\mathbb{I}_1\otimes\mathbb{I}_2$ operator contributes to the transition. Therefore, for  $\Lambda N$ final states, the  transition will match the following structure
\begin{align}
g_{1}^{\Lambda, \eta}\langle\tfrac{1}{2}\,t_{Nf}\vert\tfrac{1}{2}\,t_{Ni}\rangle\langle 0\,t_{\Xi}-\tfrac{1}{2}\vert\tfrac{1}{2}\,t_{\Xi},\tfrac{1}{2}\,-\tfrac{1}{2}\rangle.
\end{align}
%
%
The Clebsch-Gordan coefficient will only be non-zero when the initial state contains a $\Xi^0$ hyperon. By matching the former expression with that for the $\Xi^0 n \to \Lambda n$ transition, one obtains the corresponding relation between the isospin coupling and the product of weak and strong coupling constants. Thus, one has:
\begin{eqnarray}
&&g_{1}^{\Lambda, \eta}=\sqrt{2}\,g_{\Xi^0\Lambda\eta}^{W}\,g^S_{NN\eta} \ , \\
&&g_{2}^{\Lambda, \eta}=0 \ , \\
&&g_{3}^{\Lambda, \eta}=0 \ .
\end{eqnarray}
and similarly for the $\omega$ meson.

Considering now the case of  $\Sigma$N final states, only the ${O}_2^\Sigma$ operator can act in isoscalar meson exchange. Hence, following similar steps as in the previous case but for the $\Xi^0 n \to \Sigma^0 n$ transition, we find: 
\begin{eqnarray}
&&g_{1}^{\Sigma, \eta}=0,\\
&&g_{2}^{\Sigma, \eta}=\,\sqrt{2} \, g^{W}_{\Xi^0\Sigma^0\eta}\,g^S_{NN\eta},\\
&&g_{3}^{\Sigma, \eta}=0.
\end{eqnarray}

Let us now turn to the isovector ($\pi$ or $\rho$) mesons.
For $\Lambda N$ final states the first isospin structure does not contribute because the isovector meson cannot connect an isospin 0 initial state with the final $\Lambda$ at the weak vertex. The combined analysis of the $\Xi^0 p \to \Lambda p$, $\Xi^0 n \to \Lambda n$ and $\Xi^- p \to \Lambda n$ amplitudes therefore determines:
\begin{eqnarray}
&&g_{1}^{\Lambda, \pi}=0,\\
&&g_{2}^{\Lambda, \pi}=-\frac{1}{\sqrt{2}}\,g_{\Xi^-\Lambda\pi^-}^{W}\,g_{n p\pi^- }^{S}=-g_{\Xi^-\Lambda\pi^-}^{W}g_{NN\pi}^{S}  \ , \\
&&g_{3}^{\Lambda, \pi}=0 \ ,
\end{eqnarray}
where the generic strong coupling $g_{NN\pi }^S=g_{pp\pi^0 }^S=-g_{ nn\pi^0}^S=g_{np\pi^- }^S/\sqrt{2}$ has been employed in the last two terms.
Similarly, for the $\Sigma N$ final states one finds:
\begin{eqnarray}
&&g_{1}^{\Sigma,\pi}=\sqrt{2}\,g^{W}_{\Xi^0\Sigma^0\pi^0}\,g^S_{NN\pi },\\
&&g_{2}^{\Sigma, \pi}=0,\\
&&g_{3}^{\Sigma, \pi}=g^{W}_{\Xi^-\Sigma^-\pi^0}\,g^S_{NN\pi } \ .
\end{eqnarray}

Finally, the case of the isodoublet ($K$ or $K^*$) mesons involves the exchange of an isospin $\tfrac{1}{2}$ particle. After including the isospurion, as seen in Eq.~(\ref{eq:isomat}), the isospin conserving transitions are then mediated by isoscalar or isovector operators. 
In the case of $\Lambda N$ final states, working out the $\Xi^0 p \to \Lambda p$ and $\Xi^0 n \to \Lambda n$ amplitudes (and the $\Xi^- p \to \Lambda n$ one as a consistency check), we find:
\begin{eqnarray}
&&g_{1}^{\Lambda, K}=\frac{1}{\sqrt{2}}\,g_{\Xi\Lambda K}^S\left(2g_{pp{K}^0}^{W}+g_{pnK^+}^{W}\right),\\
&&g_{2}^{\Lambda, K}=-\frac{1}{\sqrt{2}}\,g_{\Xi\Lambda K}^Sg_{npK^+}^{W} \ , \\
&&g_{3}^{\Lambda, K}=0 \ ,
\end{eqnarray}
written in terms of  the generic strong coupling $g_{\Xi\Lambda {K}}^S=g_{\Xi^0\Lambda {K}^0}^S=g_{\Xi^-\Lambda K^+}^S$.

Other weak processes are possible in the case of $K$-exchange when the weak vertex is the $\Xi N K$ one. Those processes involve an interchange of particles either in the initial or final state of the amplitude but the operators mediating the transition are the same as before. Therefore, after analyzing the $p \Xi^0 \to \Lambda p$, $n \Xi^0 \to \Lambda n $ and $p \Xi^- \to \Lambda n$ amplitudes, we find: 
\begin{eqnarray}
&&\tilde{g}_{1}^{\Lambda, K}=-\frac{1}{\sqrt{2}}g_{\Lambda NK}^S\left(g_{\Xi^0 n K^0}^{W}-2g_{\Xi^0 p K^+}^{W}\right),\\
&&\tilde{g}_{2}^{\Lambda, K}=\frac{1}{\sqrt{2}}g_{\Lambda NK}^S g_{\Xi^0  n K^0}^{W} \ , \\
&&\tilde{g}_{3}^{\Lambda, K}=0 \ ,
\end{eqnarray}
written in terms of the generic strong coupling $g_{\Lambda N K}^S=g_{\Lambda p K^+}^S=g_{\Lambda n K^0}^S$.

For the $\Sigma$N final state, we find
\begin{eqnarray}
&&g_{1}^{\Sigma, K}=-\frac{1}{\sqrt{2}}g^S_{\Xi\Sigma{K}}g^{W}_{pnK^+},\\
&&g_{2}^{\Sigma, K}=\frac{1}{\sqrt{2}}g^S_{\Xi\Sigma{K}}\left(g^{W}_{pnK^+}+2g^{W}_{pp{K}^0}\right),\\
&&g_{3}^{\Sigma, K}=\frac{1}{\sqrt{2}}g^S_{\Xi\Sigma{K}}g^{W}_{pnK^+} \ ,
\end{eqnarray}
written in terms of the generic strong coupling $g^S_{\Xi\Sigma{K}}=g_{\Xi^0 \Sigma^0 {K}^0}^S=-g_{\Xi^- \Sigma^- {K}^0}^S/\sqrt{2}=-g_{\Xi^- \Sigma^0 {K}^+}^S$.

For the processes in which the weak vertex is the $\Xi N K$ one, which involve an interchange of particles in the initial or final state, we find:  
\begin{eqnarray}
&&\tilde{g}_{1}^{\Sigma, K}=\frac{1}{\sqrt{2}}\,g^S_{\Sigma N {K}}\,g^{W}_{\Xi^0 n \overline{K}^0},\\
&&\tilde{g}_{2}^{\Sigma, K}=-\frac{1}{\sqrt{2}}\,g^S_{\Sigma N {K}}\left(g^{W}_{\Xi^0 n \overline{K}^0}-2g^{W}_{\Xi^0 p K^-}\right),\\
&&\tilde{g}_{3}^{\Sigma, K}=-\frac{1}{\sqrt{2}}\,g^S_{\Sigma N {K}}\,g^{W}_{\Xi^0 n \overline{K}^0},
\end{eqnarray}
written in terms of the generic strong coupling $g^S_{\Sigma N {K}}=g^S_{p \Sigma^+  {K}^-}=-g^S_{n \Sigma^0 \overline{K}^0}= g^S_{p \Sigma^+ \overline{K}^0}/\sqrt{2}=g^S_{n \Sigma^- {K}^-}/\sqrt{2}$.

In summary, in the expression of Eq.~(\ref{eq:pseudo}) for the potential mediated by pseudoscalar mesons, 
the constants $A^{Y, \phi}_{k}$ and $B^{Y, \phi}_{k}$ correspond to the $g^{Y, \phi}_{k}$ coefficients just derived, which contain products of weak and strong couplings. The weak coupling constants employed should be the parity violating ones in the case of  the $A^{Y, \phi}_{k}$ constants and the parity conserving ones in the case of the $B^{Y, \phi}_{k}$ ones. 

Similarly, the $A^{Y, v}_{k}$, $B^{Y, v}_{k}$ and $C^{Y, v}_{k}$ couplings appearing in the vector meson exchange potential of Eq.~(\ref{eq:vector}), correspond to the $g^{Y, v}_{k}$ coefficients, but taking into account the following considerations:
\begin{itemize}  
\item $A^{Y, v}_{k}$ contains the parity violating weak coupling constant times the sum of both strong vector and tensor ones.
\item $B^{Y, v}_{k}$ contains the parity conserving weak vector coupling constant times the strong vector one.
\item $C^{Y, v}_{k}$ contains the sum of both parity conserving weak vector and tensor coupling constants times the sum of both strong vector and tensor ones.
\end{itemize}

The coupling constants are derived in the appendix and their explicit values can be found in the tables listed there.

\section{\label{sec:results}Results}
The results for the non mesonic $\Lambda$-induced, $\Lambda \Lambda \to YN$, decay rate of \sllhe{} are displayed in Tables~\ref{t:rateLLLL_no} to \ref{t:rateLLXN} and are given in terms of the free decay rate for the $\Lambda$ hyperon, $\Gamma_\Lambda=3.8\e{9}\,s^{-1}$. The possible final states are $\Lambda n$, $\Sigma^0 n$ and $\Sigma^- p$ but, by virtue of the $\Delta I =1/2$ rule, isospin coupling algebra relates the decay rates of the two later modes by a factor of 2. Therefore, the $\Lambda\Lambda \to \Sigma^0 n$ and $\Lambda\Lambda \to \Sigma^- p$ channels fulfil $\Gamma_{\Sigma^0 n}=\Gamma_{\Sigma N}/3$ and  $\Gamma_{\Sigma^- p}=2\Gamma_{\Sigma N}/3$, respectively, where $\Gamma_{\Sigma N}$ collects the total $\Sigma N$ decay rate, which is the one quoted in the Tables. We also give the contribution of each individual meson separately, in order to asses its importance in a given transition, and we add up the contribution of the lightest pseudoscalar mesons sequentially for a better interpretation of our results. 

\begin{table}[ht]
\caption{\label{t:rateLLLL_no}Individual and combined meson-exchange contributions to the nonmesonic decay rate of \sllhe{}, when only the diagonal $\Lambda\Lambda - \Lambda\Lambda$ component of the initial w.f. is included, and in the absence of final state interactions. Results given in units of $\Gamma_\Lambda=3.8\e{9}\,s^{-1}$.}
\begin{ruledtabular}
\begin{tabular}{ccc}
\textbf{Meson}&	$\bm{\Lambda}$\textbf{n}	&	$\bm{\Sigma}$\textbf{N}				\\ \hline
$\bm{\pi}$	&	-		&	$1.94\e{-2}$	\\
$\bm{K}$	&	$1.45\e{-3}$		&	$2.41\e{-3}$	\\
$\bm{\eta}$	&	$1.67\e{-4}$		&	-	\\
$\bm{\rho}$	&	-	&	$2.20\e{-3}$	\\
$\bm{K^*}$	&	$3.32\e{-4}$	&	$2.73\e{-3}$	\\
$\bm{\omega}$	&	$3.20\e{-4}$		&	-	\\
$\bm{\pi+K}$	&	$1.45\e{-3}$		&	$2.69\e{-2}$	\\
$\bm{\pi+K+\eta}$	&	$1.72\e{-3}$	&	$2.69\e{-2}$	\\
\textbf{All}	&	$2.39\e{-3}$ &	$4.22\e{-2}$	
\end{tabular}
\end{ruledtabular}
\end{table}
%

We start by presenting in Table\,\ref{t:rateLLLL_no} the contribution to the  $\Lambda \Lambda \to YN$ decay coming from the diagonal $\Lambda\Lambda-\Lambda\Lambda$ component of the wave function without the inclusion of final state interactions, followed by the results of Table\,\ref{t:rateLLLL} where these effects are considered. The strong coupling constants required to describe the $\Lambda\Lambda \to \Lambda\Lambda \to YN$ transition are taken from the Nijmegen soft-core NSC97f model \cite{Stoks:1999bz}, which has been proved to reproduce satisfactorily the scarce $YN$ scattering data, as well as the structure of $\Lambda$-hypernuclei and their decay properties. Consequently, the results presented in Tables\,\ref{t:rateLLLL_no} and \ref{t:rateLLLL} are obtained following the same approach as that of Ref.~\cite{Parreno:2001xv}, except for minor changes in the initial $\Lambda\Lambda$  and the final $YN$ wave functions, which have been obtained with higher precision here. Therefore, they have to be considered as benchmark results against which we can later asses the importance of the new $\Lambda\Lambda \to \Xi N \to YN$ transition explored in the present work.

Isospin conservation at the strong vertex excludes the exchange of a $\pi$ or a $\rho$ meson in the $\Lambda \Lambda \to \Lambda n$  transition presented in Table\,\ref{t:rateLLLL_no}, which ignores final strong interaction effects, and this is reflected as a null contribution to the decay rate for those mesons. Instead, we find that the dominant contribution to this decay mode is coming from $K$ exchange, with a  rate of $1.45\e{-3} \Gamma_\Lambda $, corresponding to roughly 60\% of the rate obtained when all mesons are considered ($\Gamma_{\Lambda n}=2.39 \e{-3} \Gamma_\Lambda $). The remaining 40\% of the $\Gamma_{\Lambda n}$ rate originates from the exchange of $K^*$, $\omega$ and $\eta$ mesons, with individual contributions which are one order of magnitude smaller than that for $K$ exchange. Conversely, for $\Sigma N$ final states, one sees a clear dominance of  the $\pi$ meson contribution, giving practically half of the total $\Sigma N$ rate of $\Gamma_{\Sigma N}=4.22\e{-2} \Gamma_\Lambda $. In this case, and also due to isospin considerations, the isoscalar $\eta$ and $\omega$ mesons do not contribute, so the remaining rate is provided by the $K$, $\rho$ and $K^*$ mesons with similar contributions, and again, one order of magnitude smaller. Adding the partial decay rates of the $\Lambda n$, $\Sigma^0n$ and $\Sigma^- p$ final states,  one obtains a total $\Lambda\Lambda \to YN$ decay rate of $\Gamma_{YN}=4.46\e{-2} \Gamma_\Lambda $, distributed into an almost negligible $\Lambda n$ contribution ($\sim5\%$ of $\Gamma_{YN}$) in front of the $\Sigma N$ one ($\sim95\%$ of $\Gamma_{YN}$).
\begin{table}[ht]
\caption{\label{t:rateLLLL}Individual and combined meson-exchange contributions to the nonmesonic decay rate of \sllhe{}, when only the diagonal $\Lambda\Lambda - \Lambda\Lambda$ component of the initial w.f. is included, and considering final state interactions. Results given in units of $\Gamma_\Lambda=3.8\e{9}\,s^{-1}$.}
\begin{ruledtabular}
\begin{tabular}{ccc}
\textbf{Meson}&	$\bm{\Lambda}$\textbf{n}	&	$\bm{\Sigma}$\textbf{N}				\\ \hline
$\bm{\pi}$	&	$1.35\e{-4}$	&		$7.15\e{-3}$	\\
$\bm{K}$	&	$2.22\e{-2}$		&	$9.69\e{-4}$	\\
$\bm{\eta}$	&	$8.95\e{-4}$		&	$7.67\e{-7}$	\\
$\bm{\rho}$	&	$1.32\e{-5}$	&	$2.37\e{-6}$	\\
$\bm{K^*}$	&	$4.28\e{-3}$	&	$2.35\e{-4}$	\\
$\bm{\omega}$	&	$4.95\e{-5}$		&	$1.42\e{-7}$	\\
$\bm{\pi+K}$	&	$2.17\e{-2}$		&	$6.34\e{-3}$	\\
$\bm{\pi+K+\eta}$	&	$1.41\e{-2}$		&	$6.29\e{-3}$	\\
\textbf{All}	&	$3.00\e{-2}$	&	$5.81\e{-3}$	
\end{tabular}
\end{ruledtabular}
\end{table}
%

As mentioned before, the results of Table\,\ref{t:rateLLLL} also correspond to the contributions to the rate from the diagonal $\Lambda\Lambda-\Lambda\Lambda$ component of the wave function, but  incorporate the effect of final state interactions, which, as can be seen, reduce the total $\Lambda \Lambda \to YN$ decay rate by about 20\% to a value  $\Gamma_{YN}=3.58\e{-2} \Gamma_\Lambda$.
Of special note is the contribution of $K$ exchange to the $\Lambda n$ mode, which gets enhanced by one order of magnitude when final state interactions are implemented, becoming the dominant mechanism for the transition. Note also that a similar enhancement is seen for the contribution of the $K^*$ meson, which represents the second dominant contribution, yet, one order of magnitude smaller than its pseudoscalar partner. In the case of $\Sigma N$ final states, we observe that final state interactions cause a  similar reduction, of about a factor 2.5, for the pseudoscalar $\pi$ and $K$ contributions, while the reduction is even larger for $\rho$ and $K^*$ vector meson exchange, giving rise to an overall decrease of the $\Sigma N$ rate by almost one order of magnitude. Consequently, the inclusion of final state interactions has inverted the relative importance of the decay modes, from 5\% to 84\% for the $\Lambda\Lambda\to\Lambda n$ channel and from 95\% to 16\% for the $\Lambda\Lambda\to\Sigma N$ one, increasing the value of the $\Gamma_{\Lambda n} / (\Gamma_{\Sigma^0 n} + \Gamma_{\Sigma^- p})$ ratio by more than a factor 90, from $0.06$ to $5.16$. Since these decay channels could, in principle, be detected separately in experiments, this ratio could be used to learn about the weak decay mechanism in the strangeness $S=-2$ sector and the role played by the strong interaction in the decay process.

Another change associated to the effect of final state interactions that can be inferred from Table\,\ref{t:rateLLLL} is that previously excluded meson exchanges now contribute, albeit in a very moderate manner. This is the case of the $\pi$ meson, for example, which now contributes to the $\Lambda \Lambda \to \Lambda n$ decay rate through the intermediate weak $\Lambda \Lambda \to \Sigma N$ transition followed by the $\Sigma N \to \Lambda n$ strong one.
%

Up to this point, our results are totally in line with those found in Ref.~\cite{Parreno:2001xv}, as expected, since the only essential difference here is 
the use of slightly different correlated baryon-baryon wave functions.
The novelty of the present paper is the consideration of the strong non-diagonal $\Lambda\Lambda -\Xi N$ mixing of the $\Lambda\Lambda$ wave function. However, contrary to the previous case, the new strong coupling constants required for the description of the $\Xi N \to YN$ transition do not have an experimental support. For this reason, we will compare the results obtained when these additional coupling constants are taken either from the same NSC97f model employed in the description of the diagonal $\Lambda\Lambda \to \Lambda\Lambda \to YN$ transition or from the chiral Lagrangians given in the appendix.

When the $\Lambda\Lambda \to \Xi N \to YN$ component is added to the calculations using the strong coupling constants given by the Nijmegen soft-core NSC97f model \cite{Stoks:1999bz}, we obtain the results of Table\,\ref{t:rateLLXN_NSC97f}.  We observe that the only significant effect of the $\Lambda\Lambda-\Xi N$ mixing to the decay into a final $\Lambda n$ state comes from $\pi$ and $K$ exchanges. Their combined effect ends up decreasing the decay rate by more than a factor two, from  $\Gamma_{\Lambda n}=3.00\e{-2}\Gamma_\Lambda$ to $1.31\e{-2}\Gamma_\Lambda$. The $\Sigma N$ decay channel experiences an increase of over a factor four, from  $\Gamma_{\Sigma N}=5.81\e{-3}\Gamma_\Lambda$ to $2.67\e{-2}\Gamma_\Lambda$. Altogether, the $\Lambda\Lambda -\Xi N$ component of the wave function brings the value of the  
$\Gamma_{\Lambda n} / (\Gamma_{\Sigma^0 n} + \Gamma_{\Sigma^- p})$ ratio to  $0.49$, a factor 10 times smaller than that found when this mixing is neglected. 
Adding the $\Lambda n$, $\Sigma^0 n$ and $\Sigma^- p$ partial rates, the total $\Lambda \Lambda \to YN$ decay rate amounts to $\Gamma_{YN}=3.98\e{-2}\Gamma_\Lambda$, which represents a modest increase of around 10\% over the case that ignored the $\Lambda\Lambda - \Xi N$ piece in the initial wave function.

In order to assess the model dependence of the $\Lambda\Lambda - \Xi N$ mixing to the $\Lambda\Lambda\to YN$ decay rate, we perform another calculation which keeps the experimentally constrained NSC97f coupling constants in the description of the weak $\Lambda\Lambda\to YN$ transition but employs, for the $\Xi N\to YN$ one,  the decay model developed in this work,  based on effective lagrangians, which is described in the appendix. The results of this hybrid model are presented in Table\,\ref{t:rateLLXN}. We observe that the addition of the partial rates yields a total contribution from the $\Lambda \Lambda \to YN$ decay mode of $\Gamma_{YN}=2.55\e{-2}\Gamma_\Lambda$, which represents an overall decrease of around 30\% over the rate obtained for the diagonal $\Lambda\Lambda$ channel only.
Comparing the results of this hybrid model with those of Table\,\ref{t:rateLLXN_NSC97f}, obtained with the strong NSC97f coupling constants, we observe a drastic reduction of almost a factor five in the $\Lambda n$ rate. This comes from the reduction by about a factor of two in the $K$-exchange rate, together with the enhancement of the $\pi$- and $\eta$-exchange contributions, with which the $K$-exchange one interferes destructively. The  $\Sigma N$ rate of the hybrid model is only 15\% smaller than that of the model employing the strong NSC97f coupling constants.
Within the hybrid model we can see that the interferences between the various meson-exchange contributions are such that the final decay rate for the $\Lambda\Lambda\to\Lambda n$ channel decreases a whole order of magnitude with respect to the case that ignores the $\Lambda\Lambda - \Xi N$ mixing, from  $\Gamma_{\Lambda n}=3.00\e{-2}\Gamma_\Lambda$ to $2.86\e{-3}\Gamma_\Lambda$. This is partially compensated by a major increase in the $\Sigma N$ decay, from  $\Gamma_{\Sigma N}=5.81\e{-3}\Gamma_\Lambda$ to $2.27\e{-2}\Gamma_\Lambda$. Altogether, the $\Lambda\Lambda -\Xi N$ component of the wave function reduces the value of the  
$\Gamma_{\Lambda n} / (\Gamma_{\Sigma^0 n} + \Gamma_{\Sigma^- p})$ ratio obtained with only the diagonal $\Lambda\Lambda \to \Lambda\Lambda$ component by a factor of 40, down to a value of $0.13$, further highlighting the effect of the $\Lambda\Lambda -\Xi N$ mixing in inverting the dominance with regards to the $\Lambda n$/$\Sigma N$ decay modes.

The effect of the $\Lambda\Lambda -\Xi N$ mixing to the $\Lambda\Lambda \to YN$ decay modes of \sllhe{} is summarized in Table\,\ref{tab:summary}, where we observe that, even if it induces a small component in the wave function, this mixing can modify moderately the rate, either by increasing it in about 10\% (NSC97f model) or by decreasing it in about 30\% (hybrid model). A more substantial change is observed in the relative importance between the $\Lambda n$ and $\Sigma N$ decay rates, which is inverted drastically, from a factor five in the absence of the $\Lambda\Lambda -\Xi N$ mixing to about 0.5 (NSC97f model) or 0.1 (hybrid model) when this new wave function component is considered. An exclusive measurement of the decay of 
\sllhe{} hypernuclei into $\Lambda n$ and $\Sigma^-p$ final states would provide valuable information to confirm the importance of the strong interaction mixing effects in the decay mechanism, and could possibly help constraining some of the strong coupling constants involving a $\Xi$ hyperon.

%
\begin{table}[ht]
\caption{\label{t:rateLLXN_NSC97f}Individual and combined meson-exchange contributions to the nonmesonic decay rate of \sllhe{}, considering final state interactions and both components of the initial w.f., $\Lambda\Lambda - \Lambda\Lambda$ and $\Lambda\Lambda - \Xi N$. The results, given in units of $\Gamma_\Lambda=3.8\e{9}\,s^{-1}$, have been obtained using the strong NSC97f model.}
\begin{ruledtabular}
\begin{tabular}{ccc}
\textbf{Meson}&	$\bm{\Lambda}$\textbf{n}	&	$\bm{\Sigma}$\textbf{N}				\\ \hline
$\bm{\pi}$	&	$3.65\e{-4}$		&	$5.85\e{-3}$	\\
$\bm{K}$	&	$1.13\e{-2}$		&	$1.37\e{-2}$	\\
$\bm{\eta}$	&	$8.62\e{-4}$		&	$1.41\e{-4}$	\\
$\bm{\rho}$	&	$1.31\e{-5}$		&	$1.86\e{-6}$	\\
$\bm{K^*}$	&	$4.27\e{-3}$		&	$2.31\e{-4}$	\\
$\bm{\omega}$	&	$4.85\e{-5}$		&	$1.36\e{-7}$	\\
$\bm{\pi+K}$	&	$7.87\e{-3}$		&	$1.97\e{-2}$	\\
$\bm{\pi+K+\eta}$	&	$3.66\e{-3}$&	$2.28\e{-2}$	\\
\textbf{All}	&	$1.31\e{-2}$	&	$2.67\e{-2}$	
\end{tabular}
\end{ruledtabular}
\end{table}
%
\begin{table}[ht]
\caption{\label{t:rateLLXN}Individual and combined meson-exchange contributions to the nonmesonic decay rate of \sllhe{}, considering final state interactions and both $\Lambda\Lambda - \Lambda\Lambda$ and $\Lambda\Lambda - \Xi N$ components of the initial w.f. The results, given in units of $\Gamma_\Lambda=3.8\e{9}\,s^{-1}$, have been obtained using the hybrid model discussed in the appendix.}
\begin{ruledtabular}
\begin{tabular}{ccc}
\textbf{Meson}&	$\bm{\Lambda}$\textbf{n}	&	$\bm{\Sigma}$\textbf{N}				\\ \hline
$\bm{\pi}$	&	$9.73\e{-4}$		&	$5.87\e{-3}$	\\
$\bm{K}$	&	$5.15\e{-3}$	&	$9.54\e{-3}$	\\
$\bm{\eta}$	&	$2.08\e{-3}$	&	$1.77\e{-4}$	\\
$\bm{\rho}$	&	$1.31\e{-5}$		&	$1.95\e{-6}$	\\
$\bm{K^*}$	&	$4.27\e{-3}$	&	$2.32\e{-4}$	\\
$\bm{\omega}$	&	$4.85\e{-5}$		&	$1.36\e{-7}$	\\
$\bm{\pi+K}$	&	$1.86\e{-3}$		&	$1.63\e{-2}$	\\
$\bm{\pi+K+\eta}$	&	$3.75\e{-4}$	&	$1.93\e{-2}$	\\
\textbf{All}	&	$2.86\e{-3}$&	$2.27\e{-2}$	
\end{tabular}
\end{ruledtabular}
\end{table}

\begin{table}
\caption{Total $\Lambda\Lambda \to YN$ contribution to the weak decay rate of \sllhe{} and ratio $\Gamma_{\Lambda n} / (\Gamma_{\Sigma^0 n} + \Gamma_{\Sigma^- p})$, considering only the diagonal component of the $\Lambda\Lambda$ wave function and including also the $\Lambda\Lambda-\Xi N$ mixing employing two different models.  The rates are given units of $\Gamma_\Lambda=3.8\e{9}\,s^{-1}$.}
\begin{ruledtabular}
\begin{tabular}{lcc}
~~~Model & $\Gamma_{YN}$ & $\Gamma_{\Lambda n} / \Gamma_{\Sigma N}$ \\
\hline
\\
$\Lambda\Lambda\to\Lambda\Lambda \to YN$ & $3.58\e{-2}$ & 5.2 \\
 & & \\
$\Lambda\Lambda\to\Lambda\Lambda \to YN$ &  &  \\
$+\Lambda\Lambda\to\Xi N \to YN$ (NSC97f) & $3.98\e{-2}$ & 0.49 \\
& & \\
$\Lambda\Lambda\to\Lambda\Lambda \to YN$ & & \\
$+\Lambda\Lambda\to\Xi N \to YN$ (hybrid) &$2.55\e{-2}$ & 0.13 
\end{tabular}
\end{ruledtabular}
\label{tab:summary}
\end{table}

The complete two-body non-mesonic decay rate $\Gamma$ of \sllhe{}, contains also the processes induced by a $\Lambda N$ pair and as such one may write $\Gamma=\Gamma_{\Lambda N\to NN}+\Gamma_{\Lambda\Lambda\to YN}$. The decay rate for the $\Lambda N\to NN$ channel has been computed\cite{Parreno:2001xv} to be $\Gamma_{\Lambda N\to NN}=0.96 \Gamma_{\Lambda}\approx 2\Gamma\left(^5_\Lambda\mathrm{He}\right)$. Comparing this result to those of the ${\Lambda\Lambda}$ induced mode calculated in the present work, one can see that the decay rate for the $\Lambda\Lambda\to YN$ transition with $\Lambda\Lambda-\Lambda\Lambda$ diagonal correlations amounts to a 3.7\% of the one-nucleon induced rate $\Gamma_{\Lambda N\to NN}$, while the inclusion of the $\Lambda\Lambda -\Xi N$ mixing produces a slight increase  in this percentage up to 4.1\% (NSC97f model) or a decrease down to 2.6\% (hybrid model).

\section{
\label{sec:conclusions}
Summary and Conclusions}

We have quantified the effects of the strong interaction in the decay rate of the \sllhe{} hypernucleus, paying special attention to the new $\Xi N \to \Lambda N$ and $\Xi N \to \Sigma N$ weak decay channels, which appear with the opening of the strong coupled state $\Lambda \Lambda \to \Xi N$. The other unexplored weak decay channels, $\Sigma \Sigma \to \Lambda N$ and $\Sigma \Sigma \to \Sigma N$, have not been addressed in the present paper due to the comparatively smallness of the $\Lambda \Lambda \to \Sigma \Sigma$ component of the initial wave function in the relevant interaction range. The new wave functions have been obtained by solving the in-medium scattering matrix ($G-$matrix) for the interacting baryons in the initial hypernucleus. In addition, the effects of the strong interaction on the final state have been studied through the solution of the scattering matrix ($T-$matrix) describing only the interaction between the two weakly emitted baryons. Our weak interaction model is based in the exchange of mesons belonging to the ground-state pseudoscalar and vector mesons, and requires the use of flavor-SU(3) and flavor-SU(3) $\times$ spin-SU(2) symmetry to determine the unknown baryon-baryon-meson coupling constants.

Our work shows remarkable sensitivity of the decay mechanism to the strong interaction. In particular, the consideration of the mixing to $\Xi N$  initial states increases the $\Lambda\Lambda$ induced decay rate by about 10\% in the case of a model that employs the NSC97f strong coupling constants or decreases it by about  30\% if the strong baryon-baryon-meson coupling constants involving a $\Xi$ hyperon are derived from a chiral effective lagrangian. 
The new $\Lambda\Lambda$ induced decay rate represents about 3-4\% of the dominant one-nucleon induced rate $\Gamma_{\Lambda N\to NN}$.

Despite the small overall contribution of the $\Lambda \Lambda$ channel to the decay of \sllhe{}, substantial changes are observed in the $\Gamma_{\Lambda n} / (\Gamma_{\Sigma^0 n} + \Gamma_{\Sigma^- p})$ ratio when the strong interaction is carefully treated. When a $\Lambda\Lambda$ correlated wave function which includes the mixing to $\Xi N$ states is employed, the relative contribution of the $\Lambda n$ and $\Sigma N$ decay rates gets inverted with respect to what is found when only $\Lambda\Lambda$ diagonal components are considered,  changing the ratio $\Gamma_{\Lambda n} / (\Gamma_{\Sigma^0 n} + \Gamma_{\Sigma^- p})$ from a value of $5$ to 0.5 or 0.1, for the two above-mentioned models of the strong $\Xi$-hyperon couplings.  This sensitivity can be used experimentally to learn about the strong interaction in the strangeness $S=-2$ sector.


\begin{acknowledgments}

This work is partly supported by the Spanish Ministerio de Economia y Competitividad (MINECO) under the project MDM-2014-0369 of ICCUB (Unidad de Excelencia 'Mar\'\i a de Maeztu'), 
and, with additional European FEDER funds, under the contracts FIS2014-54762-P and
FIS2017-87534-P.
Support has also been received from the Ge\-ne\-ra\-li\-tat de Catalunya contract 2014SGR-401 and the Secretariat for Universities and Research of the
Ministry of Business and Knowledge of the Government of Catalonia.

\end{acknowledgments}

\appendix*
\section{}

In this appendix we derive the expressions of the coupling constants at the weak and strong vertices of the diagram depicted in Fig.~\ref{fig:momenta}, employing appropriate Lagrangians. In order to give the numerical estimates of the coupling constants for the pseudoscalar mesons, we use the results of an analysis that included the weak decays of the decuplet and the octet baryons. The coupling constants for the vector mesons are obtained from a model \cite{RoxanneWeak} based on a global fit to the octet axial currents, the strong decays of the decuplet, the $s$-wave weak decays of the octet and the weak decay of the $\Omega^-$.

\subsection{\label{sec:strong} Strong baryon-baryon-meson couplings}
The description of the interaction between two baryons of the (1/2)$^+$ octet through the exchange of either a pseudoscalar or a vector meson, needs the knowledge of the interaction Lagrangian connecting two baryons and a meson for each of the vertices involved in the corresponding diagram. The formalism for the construction of such Lagrangians was developed by Callan, Coleman, Wes and Zumino in 1968~\cite{zumino,zumino2}. In these types of realizations whenever functions of the Goldstone bosons appear, they are always accompanied by at least one space-time derivative. Since the interaction with Goldstone bosons must vanish at zero momentum in the chiral limit the expansion of the Lagrangian at low energies is in powers of derivatives and pion masses.

\subsubsection{\label{sec:pseudo_strong}Pseudoscalar mesons}
The strong Lagrangian corresponding to the exchange of a pseudoscalar meson has the following form~\cite{GL84,GL85,GSS86},
\begin{align}
\mathcal{L}^S={}&\Tr\left[\overline{B}\left(i\gamma^\mu\nabla_\mu\right)B\right]-M_B\Tr\left[\overline{B}B\right]\nonumber\\&+D\Tr\left[\overline{B}\gamma^\mu\gamma_5\left\lbrace u_\mu, B\right\rbrace\right]\nonumber\\&+F\Tr\left[\overline{B}\gamma^\mu\gamma_5\left[ u_\mu, B\right]\right] ,
\label{eq:LStrong}
\end{align}
where $F=0.52$ MeV and $D=0.85$ MeV are the octet baryon to meson couplings, $B$ ($\overline{B^i_j} = (B^j_i)^\dagger \gamma_4$) is the matrix representing the inbound (outbound) baryons
\begin{align}
B=\begin{pmatrix}
\frac{1}{\sqrt{2}}\Sigma^0+\frac{1}{\sqrt{6}}\Lambda&\Sigma^{+}&p\\
\Sigma^{-}&-\frac{1}{\sqrt{2}}\Sigma^0+\frac{1}{\sqrt{6}}\Lambda&n\\
\Xi^{-}&\Xi^0&-\frac{2}{\sqrt{6}}\Lambda
\end{pmatrix},
\end{align}
and $\nabla_\mu B=\partial_\mu B+\left[\Gamma_\mu,B\right]$ the covariant derivative introduced to account for gauge invariance. The dependence on the meson fields is contained in the $\Gamma_\mu$ and $u_\mu$ operators:

\begin{gather}
\Gamma_{\mu}=\frac{1}{2}\left(u^{\dagger}\partial_{\mu}u+u\partial_{\mu}u^{\dagger}\right) \quad
u_{\mu}=\frac{i}{2}\left(u\partial_\mu u^{\dagger}-u^{\dagger}\partial_\mu u\right),
\end{gather}
where $u$ is defined as $u=\mathrm{e}^{i\frac{\phi}{\sqrt{2}\,f_\pi}}\simeq 1+i\frac{1}{\sqrt{2}f_\pi}\phi$, with $f_\pi=93$ MeV the pion decay constant, and $\phi$ the selfadjoint matrix of inbound pseudoscalar mesons, 
\begin{align}
\phi=\begin{pmatrix}
\frac{1}{\sqrt{2}}\pi^0+\frac{1}{\sqrt{6}}\eta&\pi^{+}&K^{+}\\
\pi^{-}&-\frac{1}{\sqrt{2}}\pi^0+\frac{1}{\sqrt{6}}\eta&K^0\\
K^{-}&\overline{K}^{\,0}&-\frac{2}{\sqrt{6}}\eta
\end{pmatrix}.
\end{align}

The Lagrangian of Eq.~(\ref{eq:LStrong}) allows us to derive the Yukawa-type coupling constants of the baryons to the pseudoscalar mesons displayed in Table~\ref{T:StrongPS}.

\begin{table}
\caption{\label{T:StrongPS}Strong pseudoscalar meson couplings to the octet baryons, where $D$ and $F$  are the couplings of the pseudoscalar Lagrangian of Eq.~(\ref{eq:LStrong}) and $\overline{M}$ denotes the average mass of the baryons at the baryon-baryon-meson vertex.}
\begin{ruledtabular}
\begin{tabular}{ccc}
\textbf{Coupling}&\textbf{Analytic value}&$\bm{g^S_{BB\phi}}$\\ \hline
$NN\pi$&$\frac{D+F}{2f_{\pi}}2\overline{M}$&13.83\\
$NN\eta$&$\frac{3F-D}{2\sqrt{3}f_{\pi}}2\overline{M}$&4.14\\
$\Lambda N K$&$-\frac{3D+F}{2\sqrt{3}f_{\pi}}2\overline{M}$&-15.37\\
$\Lambda\Lambda\eta$ & $-\frac{D}{\sqrt{3}f_{\pi}}2\overline{M}$ & -11.77\\
$\Lambda\Sigma\pi$ & $\frac{D}{\sqrt{3}f_{\pi}}2\overline{M}$ & 12.19\\
$\Sigma N K$&$\frac{D-F}{2f_{\pi}}2\overline{M}$&3.78\\	
$\Sigma\Sigma\pi$ & $\frac{F}{f_{\pi}}2\overline{M}$ & 13.36\\	
$\Sigma\Sigma\eta$ & $\frac{D}{\sqrt{3}f_{\pi}}2\overline{M}$ & 12.60\\	
$\Xi\Lambda K$&$\frac{3F-D}{2\sqrt{3}f_{\pi}}2\overline{M}$&5.36\\
$\Xi\Sigma K$&$-\frac{D+F}{2f_{\pi}}2\overline{M}$&-18.47\\
$\Xi\Xi\pi$ & $-\frac{D-F}{\sqrt{2}f_{\pi}}2\overline{M}$ & -4.68\\
$\Xi\Xi\eta$ & $-\frac{3F+D}{2\sqrt{3}f_{\pi}}2\overline{M}$ & -19.77
\end{tabular}
\end{ruledtabular}
\end{table}

\subsubsection{\label{sec:vector_strong}Vector mesons}
The interaction between baryons and vector mesons has not been as extensively studied as the one involving pseudoscalar mesons, but one can use hidden local symmetry (HLS)\cite{HLS-85}, to accommodate vector mesons consistently with chiral symmetry. In order to incorporate these mesons in our formalism, the following Lagrangian is used: 
\begin{align}
\mathcal{L}_{VBB}=&-g\big\lbrace\langle\overline{B}\gamma_{\mu}[V^{\mu}_8,B]\rangle+\langle\overline{B}\gamma_{\mu}B\rangle\langle V^{\mu}_8\rangle\nonumber\\
&+\frac{1}{4M}\big(F\langle \overline{B}\sigma_{\mu\nu}[\partial^\mu V^{\nu}_8-\partial^\nu V^{\mu}_8,B]\rangle\nonumber\\
&+D\langle\overline{B}\sigma_{\mu\nu}\lbrace \partial^\mu V^{\nu}_8-\partial^\nu V^{\mu}_8,B\rbrace\rangle\big)\nonumber\\
&+\langle\overline{B}\gamma_{\mu}B\rangle\langle V^{\mu}_0\rangle+\frac{C_0}{4M}\langle\overline{B}\sigma_{\mu\nu}V^{\mu\nu}_0B\rangle\big\rbrace,
\label{eq:strong_BBV}
\end{align}
which may be obtained from the generalization of the HLS formalism in SU(2) to the SU(3) sector. There, $V_8$ and $V_0$ are the octet and singlet terms in the vector meson matrix respectively,
\begin{align}
V_\mu=\frac{1}{2}\begin{pmatrix}
\rho^0+\omega & \sqrt{2}\rho^{+} & \sqrt{2}K^{*+}\\
\sqrt{2}\rho^{-} & -\rho^0+\omega & \sqrt{2}K^{*0}\\
\sqrt{2}K^{*-} & \sqrt{2}\overline{K}^{\,*0} & \sqrt{2}\phi
\end{pmatrix}_{\mu} \, ,
\end{align}
the SU(3) $D$ and $F$ constants take now the values $D=2.4$ and $F=0.82$, and the constant $C_0$ is chosen to be $3F-D$, such that the $\phi NN$ vertex is null (according to naive expectations based in the Okubo  Zweig  Iizuka, {\it OZI}, rule) and the anomalous magnetic coupling of the $\omega NN$ vertex gives $\kappa_\omega\simeq 3F-D$\cite{khanchan}. The baryon mass is represented by $M$, while $g$ takes the form:
\begin{align}
g=\frac{m}{\sqrt{2}f_\pi},
\end{align}
where $m$ is the mass of the exchanged meson.

The octet and singlet matrices can be obtained by considering the mixing of the octet and singlet components of the physical $\omega$ and $\phi$ meson, which under the ideal mixing assumption leads to\cite{okubo63}:
\begin{gather}
\omega=\sqrt{\frac{1}{3}}\omega_8+\sqrt{\frac{2}{3}}\omega_0,\\
\phi=-\sqrt{\frac{2}{3}}\phi_8+\sqrt{\frac{1}{3}}\phi_0.
\end{gather}

The Yukawa couplings involving vector mesons are displayed in Table~\ref{T:StrongV}.

\begin{table}
\caption{\label{T:StrongV}Strong vector meson couplings to the octet baryons, with $g$, $D$, $F$ and $C_0$ being the couplings of the strong $VBB$ Lagrangian of Eq.~\ref{eq:strong_BBV}.}
\begin{ruledtabular}
\begin{tabular}{ccc}
\textbf{Coupling}&\textbf{Analytic value}&$\bm{g^S_{BBV}}$\\ \hline
$NN\rho$ $(V)$ &$-\frac{1}{2}g$&2.95\\
$NN\rho$ $(T)$ &$-\frac{D+F}{2}g$&9.49\\
$NN\omega$ $(V)$ &$\frac{2\sqrt{2}+3}{2\sqrt{3}}g$&3.54\\
$NN\omega$ $(T)$ &$\frac{\sqrt{2}C_0+D+F}{2\sqrt{3}}g$&5.62\\
$\Lambda N K^*$ $(V)$ &$-\frac{\sqrt{3}}{2}g$&-5.11\\
$\Lambda N K^*$ $(T)$ &$-\frac{D+3F}{2\sqrt{3}}g$&-8.27\\
$\Lambda\Lambda\omega$ $(V)$ &$\frac{\sqrt{2}+1}{\sqrt{3}}g$&8.22\\
$\Lambda\Lambda\omega$ $(T)$ &$\frac{\sqrt{2}C_0+2D}{6\sqrt{3}}g$&2.77\\
$\Lambda\Sigma\rho$ $(V)$ &0 &0\\
$\Lambda\Sigma\rho$ $(T)$ &$\frac{D}{\sqrt{3}}g$&8.17\\
$\Sigma N K^*$ $(V)$ &$-\frac{1}{2}g$&-2.95\\
$\Sigma N K^*$ $(T)$ &$\frac{D-F}{2}g$&4.66\\
$\Sigma\Sigma\rho$ $(V)$ &$-g$&-5.89\\
$\Sigma\Sigma\rho$ $(T)$ &$-Fg$&-4.83\\
$\Sigma\Sigma\omega$ $(V)$ &$\frac{\sqrt{2}+1}{\sqrt{3}}g$&8.22\\
$\Sigma\Sigma\omega$ $(T)$ &$\frac{\sqrt{2}C_0+2D}{2\sqrt{3}}g$&8.31\\
$\Xi\Lambda K^*$ $(V)$ & $-\frac{\sqrt{3}}{2}g$&5.11\\
$\Xi\Lambda K^*$ $(T)$&$\frac{D-3F}{2\sqrt{3}}g$&0.10\\
$\Xi\Sigma K^*$ $(V)$ &$-\frac{1}{2}g$&-2.95\\
$\Xi\Sigma K^*$ $(T)$ &$-\frac{D+F}{2}g$&-9.49\\
$\Xi\Xi\rho$ $(V)$ &$\frac{1}{2}g$&2.95\\
$\Xi\Xi\rho$ $(T)$ &$-\frac{F-D}{2}g$&-4.66\\
$\Xi\Xi\omega$ $(V)$ &$\frac{2\sqrt{2}+1}{2\sqrt{3}}g$&6.51\\
$\Xi\Xi\omega$ $(T)$ &$\frac{D-F}{2\sqrt{3}}g$&2.69
\end{tabular}
\end{ruledtabular}
\end{table}

\subsection{Weak baryon-baryon-meson vertices: PV contribution}\label{ch:Weak}
\subsubsection{Pseudoscalar mesons}
The starting point to derive the weak vertices is the heavy baryon chiral perturbation Hamiltonian introduced by Jenkins and Manohar \cite{Jenkins:1990jv,RoxanneWeak} to account for strangeness changing amplitudes, all the while neglecting those terms in which the decuplet baryon matrices appear.  Using a lowest-order chiral analysis one can only generate parity violating amplitudes, since the weak chiral Lagrangian describing parity-conserving transitions has the wrong transformation property under the combined action of the charge and parity operators~\cite{MRR69}. The effective Lagrangian: 
\begin{align} 
\label{eq:Lef_feble}
\mathcal{L}={}&\sqrt{2}\big(h_D\Tr\left[\overline{B}\left\lbrace \xi^{\dagger}h\xi,B\right\rbrace\right]\nonumber\\&+h_F\Tr\left[\overline{B}\left[ \xi^{\dagger}h\xi,B\right]\right]\big),
\end{align}
is written in terms of the dimensionless constants $h_D=-1.69\e{-7}$ and $h_F=3.26\e{-7}$, which can be fitted to reproduce known meson decay amplitudes and the s-wave nonleptonic weak decays of the baryon octet members \cite{RoxanneWeak}. The \textit{h} operator is a $3\times 3$ matrix with a single non-zero element, $h_{23}=1$, which accounts for strangeness variations of $\vert\Delta S\vert=1$. The operator $\xi$ plays a role equivalent to the one of the \textit{u} operator in the strong Lagrangian defined in the previous section. The Lagrangian of Eq.~(\ref{eq:Lef_feble}) allows one to find the weak PV coupling constants of the baryons to the pseudoscalar mesons displayed in Table~\ref{T:WeakPVPS}.

\begin{table}
\caption{\label{T:WeakPVPS}Weak PV pseudoscalar meson couplings to the octet baryons.}
\begin{ruledtabular}
\begin{tabular}{ccc}
\textbf{Coupling}&\textbf{Analytic value}&$\bm{g^{PV}_{BB\phi}}$\\ \hline
%
$pnK^+$&$h_D+h_F$&$1.57\e{-7}$\\
%
$ppK^0$&$h_F-h_D$&$4.95\e{-7}$\\
%
$nnK^0$&$2h_F$&$6.52\e{-7}$\\
$\Xi^0 n \overline{K}^0$ & $0$ & $0$ \\
$\Xi^0 p K^-$&$0$ & $0$ \\
%
$\Xi^0\Lambda\pi^0$ & $\frac{3h_F-h_D}{2\sqrt{3}}$&$3.31\e{-7}$\\
%
$\Xi^{-}\Lambda\pi^-$&$\frac{h_D-3h_F}{\sqrt{6}}$&$-4.68\e{-7}$\\
%
$\Xi^0\Lambda\eta$ & $\frac{h_D-3h_F}{2}$&$-5.73\e{-7}$\\
%
$\Xi^0\Sigma^0\pi^0$ & $-\frac{h_D+h_F}{2}$ &$-7.84\e{-8}$\\
%
$\Xi^-\Sigma^-\pi^0$&$\frac{h_D+h_F}{\sqrt{2}}$&$1.11\e{-7}$\\
%
$\Xi^-\Sigma^0\pi^-$&$-\frac{h_D+h_F}{\sqrt{2}}$&$-1.11\e{-7}$\\
%
$\Xi^0\Sigma^0\eta$ & $\frac{\sqrt{3}(h_D+h_F)}{2}$ &$1.36\e{-7}$ \\
%
$\Xi^-\Sigma^-\eta$&$\frac{\sqrt{3}(h_D+h_F)}{\sqrt{2}}$&$1.92\e{-7}$
\end{tabular}
\end{ruledtabular}
\end{table}

\subsubsection{Vector mesons}
For the weak vertices the introduction of the SU(6)$_W$ group is necessary. This group describes the product of the SU(3) flavour group with the SU(2)$_W$ spin group, which is the proper group to consider when dealing with particles in motion, as the ones involved in weak decay processes \cite{Close79}. In this representation the meson fields are expressed in terms of a quark-antiquark product $\phi^a_b$, where the upper and lower indices refer to the spin-flavor antiquark and quark combinations respectively.
\begin{align}
\phi^a_b=\varepsilon q_b\overline{q}_a\text{ with}\begin{cases}
\varepsilon=1 & \text{if both a and b even}\\
\varepsilon=-1 & \text{otherwise}.
\end{cases}
\end{align}

The labels used correspond to the fundamental representation of SU(6)$_W$, and as such, both indices range from 1 to 6. The spin up and spin down $u$ quarks are assigned to 1 and 2 respectively, the $d$ quarks are assigned to 3 and 4 and the strange quarks to 5 and 6.

For the baryons one must define the symmetric tensors
\begin{gather}
B^{abc}\equiv\frac{1}{6}\sum_{\substack{perm\\a,b,c}}S^a(1)S^b(2)S^c(3)\\
B_{abc}=\overline{B}^{abc}=\frac{1}{6}\sum_{\substack{perm\\a,b,c}}\overline{S}^a(1)\overline{S}^b(2)\overline{S}^c(3),
\end{gather}
where the constants $a$, $b$ and $c$ run over the same numerical values stated before. The couplings may be found by expressing the Hamiltonian in terms of the SU(6)$_W$ tensors. This Hamiltonian is the product of two currents, each belonging to the 35 representation, and using the Clebsch-Gordan series one can extract the parity conserving and parity violating pieces of the Hamiltonian. As discussed before, imposing the right CP transformation leads to only  PV contributions, which can be expressed in terms of reduced matrix elements for the product of the appropriate representations:
\begin{gather}
2a_T:\left[(\overline{B}B)_{35}\times M_{35}\right]_{280_a}\\
2a_V:\left[(\overline{B}B)_{35}\times M_{35}\right]_{\overline{280}_a}\\
b_T:\left[(\overline{B}B)_{405}\times M_{35}\right]_{280_a}\\
b_V:\left[(\overline{B}B)_{405}\times M_{35}\right]_{\overline{280}_a}\\
c_V:\left[(\overline{B}B)_{35}\times M_{35}\right]_{35_a},
\end{gather}
with the constants $a_T$, $a_V$, $b_T$, $b_V$ and $c_V$ related to known amplitudes\cite{desplanques1980unified}:
\begin{gather}
a_T=\frac{1}{3}a_V=\frac{3}{5}G\cos\theta_c\sin\theta_c\langle\rho^0\vert V_\mu^3\vert 0\rangle\langle p\vert A^{\mu 3}\vert p\rangle,\\
b_V=-b_T=6\left(\frac{1}{\sqrt{3}}A_{\Lambda p}+A_{\Sigma^+ p}\right),\\
c_V=3\left(\sqrt{3}A_{\Lambda p}+A_{\Sigma^+ p}\right).
\end{gather}

The values $A_{\Sigma^+p}=-3.27\e{-7}$ and $A_{\Lambda p}=3.25\e{-7}$ obtained from data on the experimental angular distribution of the decay products and on the polarization of the final baryon\cite{donoghue2} are used. The final general expression accounting for the weak PV baryon-baryon-vector meson couplings is:
\scriptsize
\begin{align}
a_T\big[\overline{B}^{ij2}B_{ij1}\overline{\phi}^{6}_{3}-\overline{B}^{ij3}B_{ij6}\overline{\phi}^{1}_{2}-\overline{B}^{ij1}B_{ij2}\overline{\phi}^{5}_{4}+\overline{B}^{ij4}B_{ij5}\overline{\phi}^{2}_{1}\big]\nonumber\\
a_V\big[\overline{B}^{ij2}B_{ij5}\overline{\phi}^{2}_{3}-\overline{B}^{ij3}B_{ij2}\overline{\phi}^{5}_{2}-\overline{B}^{ij1}B_{ij6}\overline{\phi}^{1}_{4}+\overline{B}^{ij4}B_{ij1}\overline{\phi}^{6}_{1}\big]\nonumber\\
b_T\big[\overline{B}^{ij2}B_{i16}\overline{\phi}^{j}_{3}-\overline{B}^{ij3}B_{i16}\overline{\phi}^{j}_{2}-\overline{B}^{23i}B_{ij6}\overline{\phi}^{1}_{j}+\overline{B}^{23i}B_{1ij}\overline{\phi}^{6}_{j}\nonumber\\
-\overline{B}^{i1j}B_{25i}\overline{\phi}^{j}_{4}+\overline{B}^{ij4}B_{i25}\overline{\phi}^{j}_{1}+\overline{B}^{i14}B_{ij5}\overline{\phi}^{2}_{j}-\overline{B}^{i14}B_{ij2}\overline{\phi}^{5}_{j}\big]\nonumber\\
b_V\big[\overline{B}^{ij2}B_{i25}\overline{\phi}^{j}_{3}-\overline{B}^{ij3}B_{i25}\overline{\phi}^{j}_{2}+\overline{B}^{i23}B_{ij5}\overline{\phi}^{2}_{j}-\overline{B}^{23i}B_{ij2}\overline{\phi}^{5}_{j}\nonumber\\
-\overline{B}^{1ij}B_{16i}\overline{\phi}^{j}_{4}+\overline{B}^{ij4}B_{16i}\overline{\phi}^{j}_{1}-\overline{B}^{14i}B_{ij6}\overline{\phi}^{1}_{j}+\overline{B}^{i14}B_{1ij}\overline{\phi}^{6}_{j}\big]\nonumber\\
c_V\big[\overline{B}^{ijk}B_{ij6}\overline{\phi}^{k}_{4}-\overline{B}^{ij4}B_{ijk}\overline{\phi}^{6}_{k}-\overline{B}^{ijk}B_{ij5}\overline{\phi}^{k}_{3}+\overline{B}^{ij3}B_{ijk}\overline{\phi}^{5}_{k}\big] \, , 
\end{align}
\normalsize
where the SU(6)$_W$ tensor terms must be expanded in terms of the physical fields to write down the $\langle B'M\vert H_{PV}\vert B\rangle$ elements of interest. The resulting expressions and numerical results of the weak PV coupling constants of baryons to vector mesons are shown in Table~\ref{T:WeakPVV}.
%
%
%
\begin{table}
\caption{\label{T:WeakPVV}Weak PV vector meson couplings to the octet baryons.}
\begin{ruledtabular}
\begin{tabular}{ccc}
\textbf{Coupling}&\textbf{Analytic value}&$\bm{g^{PV}_{BBV}}$\\ \hline
%
%
$pnK^{*+}$ & $\frac{1}{9}\left( - b_T+ 2 b_V- 5 c_V \right) $ & $-6.72\e{-7}$ \\
$ppK^{*0}$ & $\frac{1}{9}\left( 8 a_T+ b_T-\frac{1}{2}b_V+c_V \right)$ & $1.38\e{-7}$ \\
$nnK^{*0}$ & $\frac{1}{9}\left(-2 a_T-\frac{1}{2}b_T+ b_V+ c_V \right)$ & $-4.34\e{-7}$ \\
$\Xi^0n\overline{K}^{*0}$ & $0$ & $0$ \\
$\Xi^0pK^{*-}$ & $\frac{1}{9}\left(b_T-2b_V\right)$ & $2.79\e{-7}$ \\
$\Xi^0\Lambda\rho^0$ & $\frac{\sqrt{3}}{9}\left(a_T+\frac{1}{4}b_T -\frac{1}{4}b_V+\frac{1}{2}c_V\right)$ & $1.31\e{-7}$ \\
$\Xi^- \Lambda \rho^-$ & $\frac{\sqrt{6}}{9}\left(-a_V+\frac{1}{4}b_T-\frac{1}{4}b_V+\frac{1}{2}c_V\right)$ & $2.84\e{-7}$ \\
$\Xi^0\Lambda\omega$ & $\frac{1}{3}\left( \frac{1}{3}a_T+\frac{1}{4}b_T-\frac{1}{4}b_V+\frac{1}{6}c_V\right)$ & $1.69\e{-7}$ \\
$\Xi^0\Sigma^0\rho^0$ & $\frac{1}{9}\left(-5 a_T-\frac{7}{4}b_T+\frac{5}{4}b_V-\frac{5}{2}c_V\right) $ & $-4.25\e{-7}$ \\
$\Xi^-\Sigma^-\rho^0$ & $-\frac{5\sqrt{2}}{18}\left(2a_T+c_V\right)$ & $-2.07\e{-7}$ \\
$\Xi^- \Sigma^0 \rho^-$ & $\frac{\sqrt{2}}{9}\left(-5a_V+\frac{1}{4}b_T-\frac{1}{4}b_V+\frac{5}{2}c_V\right)$ & $5.56\e{-7}$ \\
$\Xi^0\Sigma^0\omega$ & $\frac{\sqrt{3}}{27}\left( -5 a_T-\frac{1}{4}b_T-\frac{1}{4}b_V-\frac{5}{2}c_V \right) $ & $-8.47\e{-8}$ \\
$\Xi^-\Sigma^-\omega$ & $-\frac{5\sqrt{6}}{27}\left(a_T+\frac{1}{2}c_V\right)$ & $-1.20\e{-7}$ 
\end{tabular}
\end{ruledtabular}
\end{table}

\subsection{Weak baryon-baryon-meson vertices: PC contribution}
As stated above, the use of the weak effective Hamiltonian at lowest order allows us to obtain only the parity violating amplitudes. The standard method to compute the parity conserving amplitudes is based in the pole model\cite{donoghue}, according to which the weak transition is shifted from the meson vertex to the baryonic (and mesonic) line. The starting point is to consider the transition amplitude for the nonleptonic emission of a meson, $B\rightarrow B'M_i(q)$
\begin{align}
\langle B'M_i(q)\vert H_W\vert B\rangle=&\int \mathrm{d}^4x \, \mathrm{e}^{iqx}\theta(x^0)\nonumber\\
&\times\langle B'\vert [\partial A_i(x),H_W]\vert B\rangle,
\label{eq:PM_am}
\end{align}
where 
$A_i(x)$ is the axial current associated to the meson field and $H_W$ is the weak interaction Lagrangian. Inserting a complete set of states $\sum_n\vert n\rangle\langle n\vert$ into Eq.\,\eqref{eq:PM_am} leads to a series of contributions among which the dominant one corresponds to the baryon $(1/2)^+$ pole terms, that become singular in the SU(3) soft-meson limit and represent the leading contribution to the PC amplitudes \cite{donoghue2}:
\small
\begin{align}
&\langle B'M_i(q)\vert H_W\vert B\rangle\sim\nonumber\\
&\sum_n\left[\delta(\vec{p}_n-\vec{p}_{B'}-\vec{q}\,))\frac{\langle B'\vert A^{\mu}_i(0)\vert n\rangle\langle n\vert H_W(0)\vert B\rangle}{p_B^0-p_n^0}\right]+\nonumber\\
&\sum_{n'}\left[\delta(\vec{p}_B-\vec{p}_{n'}-\vec{q}\,)\frac{\langle B'\vert H_W(0)\vert n'\rangle\langle n'\vert A^{\mu}_i(0)\vert B\rangle}{p_B^0-q^0-p_{n'} ^0}\right].
\label{eq:PMNN}
\end{align}
\normalsize

\begin{figure}[!ht]
	\centering
	\includegraphics[width=0.4\textwidth]{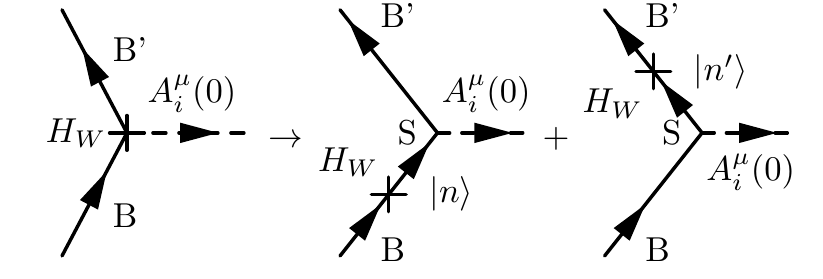}%
	\caption{\label{fig:PMfig}Pole model diagrams contributing to the weak PC conserving baryon-baryon-meson amplitudes. The label S denotes a strong BBM vertex.}
\end{figure}

For the calculation of the weak pole vertices it is necessary to express the physical states in terms of the baryon octet fields $\vert B_i\rangle$, as well as the meson states in terms of $\vert M_i\rangle$\cite{delatorre}. Furthermore, the mesonless weak transition between baryons, $\langle B\vert H_W (0)\vert B'\rangle$,  can be computed using low-energy theorems for mesons. These theorems express the matrix element for the emission of a meson of zero (or small) four-momentum in terms of the corresponding matrix element in the absence of the soft meson and some equal-time commutators of currents\cite{AD68}. They are based in the existence of certain symmetry in a given physical process, which give rise to degenerate multiplets (a state containing an arbitrary number of Goldstone bosons) with couplings related by the symmetry. Therefore, we will be able to relate the strong scattering amplitudes to the weak vertices,
\begin{align}
\lim_{q\rightarrow 0}\langle B\xrightarrow{PV}B'M_i \rangle&=\lim_{q\rightarrow 0}\langle B'M_i\vert H_{PV}\vert B\rangle\nonumber\\
&=-\frac{i}{f_{\pi}}\langle B'\vert[F_{i},H_{6}]\vert B\rangle,
\label{eq:SMRT}
\end{align}
where, following Cabibbo's theory, we have assumed that the weak hamiltonian transforms like the sixth component of an octet, $H_6$, according to the $CP$ invariance
of $H_W^{\Delta S =1}$, and $F_i$ are the corresponding 
$SU(3)$ generators.

To compute the last term of Eq.\,\eqref{eq:SMRT}, one can use the completely antisymmetric, $f_{ijk}$, and symmetric, 
$d_{ijk}$, SU(3) coefficients\cite{Grigorescu:1984hc} to
express the action of the $F_i$ generator on a baryon field,
\begin{align}
F_i\vert B_j\rangle=if_{ijk}\vert B_k\rangle ,
\label{eq:FiBario}
\end{align}
and the weak transition between baryon fields in terms of 
two reduced matrix elements, $A$ and $B$, 
\begin{align}
\langle B_k\vert H_6\vert B_j\rangle=iAf_{6jk}+Bd_{6jk} ,
\label{eq:H6}
\end{align}
which can be determined by a fit to experimental data for specific PV transitions, for which we choose the 
$\Sigma^{+}\rightarrow p+\pi^0$ ($A_{\Sigma^{+}p}$) and $\Lambda\rightarrow p+\pi^{-}$ ($A_{\Lambda p}$)
processes,
\begin{gather}
A_{\Sigma^{+}p}=\frac{i}{4f_{\pi}}\left( B-A\right),\\
A_{\Lambda p}=-\frac{i}{f_{\pi}}\frac{-3A-B}{4\sqrt{3}}.
\label{eq:ALambdaP}
\end{gather}
Combining these expressions, we obtain:
\begin{gather}
-\frac{i}{f_{\pi}}A=A_{\Sigma^{+}p}-\sqrt{3}A_{\Lambda p},\\
-\frac{i}{f_{\pi}}B=-\sqrt{3}A_{\Lambda p}-3A_{\Sigma^{+}p}.
\end{gather}
%
Therefore, when inserting the above relations in Eq.~(\ref{eq:H6}), one can obtain the weak PC baryon transitions required in the pole model,  $B_j \leftrightarrow B_k$ ,  in terms of the  $\Sigma^{+}\rightarrow p+\pi^0$ and $\Lambda\rightarrow p+\pi^{-}$ PV amplitudes.

One should note that, in principle, contributions to the PC amplitudes coming from the poles in the meson propagator are also possible. These contributions have not been included, in part due to their small contribution in comparison to those of baryon poles, but also due to the uncertainty in the phase between baryon and meson pole terms.

The expressions of the weak PC coupling constants of baryons to pseudoscalar and vector mesons are shown in Tables~\ref{taula:resultsWPC} and \ref{taula:resultsWPCVector}, respectively. The strong coupling constants in the expressions of Table~\ref{taula:resultsWPC}  should be replaced by the numerical values listed in Table~\ref{T:StrongPS}, to obtain the weak PC couplings involving pseudoscalar mesons. Analogously the weak PC vector and tensor couplings involving vector mesons are obtained by inserting, respectively, the vector and tensor values of the strong coupling constants listed in Table~\ref{T:StrongV} into the expressions of Table~\ref{taula:resultsWPCVector}.

\bibliography{references}
\begin{sidewaystable*}
\caption{\label{taula:resultsWPC}Weak baryon-baryon-pseudoscalar meson parity-conserving couplings, $g^{PC}_{BB\phi}$.}
\begin{ruledtabular}
\resizebox{\textwidth}{!}{
\begin{tabular}{ccc}
\textbf{Coupling}&\textbf{Analytic value}\\ \hline
%
$pnK^+$&$g^{S}_{\Lambda pK^+}\frac{1}{m_n-m_\Lambda}\frac{-A_{\Lambda p}}{\sqrt{2}}+g^{S}_{\Sigma^0pK^+}\frac{1}{m_n-m_{\Sigma^0}}\frac{-A_{\Sigma^+p}}{\sqrt{2}}$\\
%
$ppK^0$&$g^{S}_{p\Sigma^+\overline{K}^0}\frac{1}{m_p-m_{\Sigma^+}}A_{\Sigma^+p}$\\
%
$nnK^0$&$g^{S}_{n\Lambda\overline{K}^0}\frac{1}{m_n-m_\Lambda}\frac{-A_{\Lambda p}}{\sqrt{2}}+g^{S}_{n\Sigma^0\overline{K}^0}\frac{1}{m_n-m_{\Sigma^0}}\frac{-A_{\Sigma^+p}}{\sqrt{2}}$\\
%
$\Xi^0n\overline{K}^0$&$g^{S}_{\Lambda nK^0}\frac{1}{m_{\Xi^0}-m_{\Lambda}}\frac{1}{\sqrt{2}}\left(A_{\Lambda p}-\sqrt{3}A_{\Sigma^+p}\right) + g^{S}_{\Sigma^0 nK^0}\frac{1}{m_{\Xi^0}-m_{\Sigma^0}}\frac{-1}{2\sqrt{2}}\left(A_{\Sigma^+ p}+\sqrt{3}A_{\Lambda p}\right) + g^{S}_{\Xi^0\Lambda K^0}\frac{1}{m_{n}-m_{\Lambda}}\frac{-A_{\Lambda p}}{\sqrt{2}} +g^{S}_{\Xi^0\Sigma^0 K^0}\frac{1}{m_{n}-m_{\Sigma^0}}\frac{-A_{\Sigma^+ p}}{\sqrt{2}}$\\
%
$\Xi^0pK^{-}$&$g^{S}_{\Lambda pK^+}\frac{1}{m_{\Xi^0}-m_{\Lambda}}\frac{1}{\sqrt{2}}\left(A_{\Lambda p}-\sqrt{3}A_{\Sigma^+p}\right) +g^{S}_{\Sigma^0p K^+}\frac{1}{m_{\Xi^0}-m_{\Sigma^0}}\frac{-1}{2\sqrt{2}}\left(A_{\Sigma^+ p}+\sqrt{3}A_{\Lambda p}\right) + g^{S}_{\Xi^0\Sigma^+ K^+}\frac{1}{m_{p}-m_{\Sigma^+}}A_{\Sigma^+p}$\\
%
$\Xi^0\Lambda\pi^0$&$g^{S}_{\Xi^0\Xi^0\pi^0}\frac{1}{m_{\Lambda}-m_{\Xi^0}}\frac{1}{\sqrt{2}}\left(A_{\Lambda p}-\sqrt{3}A_{\Sigma^+p}\right)+g^{S}_{\Lambda\Sigma^0\pi^0}\frac{1}{m_{\Xi^0}-m_{\Sigma^0}}\frac{-1}{2\sqrt{2}}\left(A_{\Sigma^+ p}+\sqrt{3}A_{\Lambda p}\right)$\\
%
$\Xi^{-}\Lambda\pi^-$&$g^{S}_{\Sigma^-\Lambda\pi^+}\frac{1}{m_{\Xi^-}-m_{\Sigma^-}}\frac{1}{2}\left(\sqrt{3}A_{\Lambda p}+A_{\Sigma^+p}\right)+g^{S}_{\Xi^-\Xi^0\pi^+}\frac{1}{m_\Lambda-m_{\Xi^0}}\frac{1}{\sqrt{2}}\left(A_{\Lambda p}-\sqrt{3}A_{\Sigma^+p}\right)$\\
%
$\Xi^0\Lambda\eta$&$\left(g^{S}_{\Xi^0\Xi^0\eta}-g^{S}_{\Lambda\Lambda\eta}\right)\frac{1}{m_{\Lambda}-m_{\Xi^0}}\frac{1}{\sqrt{2}}\left(A_{\Lambda p}-\sqrt{3}A_{\Sigma^+p}\right)$\\
%
$\Xi^0\Sigma^0\pi^0$&$g^{S}_{\Xi^0\Xi^0\pi^0}\frac{1}{m_{\Sigma^0}-m_{\Xi^0}}\frac{-1}{2\sqrt{2}}\left(A_{\Sigma^+ p}+\sqrt{3}A_{\Lambda p}\right)+g^{S}_{\Sigma^0\Lambda\pi^0}\frac{1}{m_{\Xi^0}-m_{\Lambda}}\frac{1}{\sqrt{2}}\left(A_{\Lambda p}-\sqrt{3}A_{\Sigma^+p}\right)$\\
%
$\Xi^-\Sigma^-\pi^0$&$\left(g^{S}_{\Xi^-\Xi^-\pi^0}-g^{S}_{\Sigma^-\Sigma^-\pi^0}\right)\frac{1}{m_{\Sigma^-}-m_{\Xi^-}}\frac{1}{2}\left(\sqrt{3}A_{\Lambda p}+A_{\Sigma^+p}\right)$\\
%
$\Xi^-\Sigma^0\pi^-$&$\left(g^{S}_{\Sigma^-\Sigma^0\pi^+}\frac{1}{m_{\Xi^-}-m_{\Sigma^-}}+\frac{-1}{\sqrt{2}}g^{S}_{\Xi^-\Xi^0\pi^+}\frac{1}{m_{\Sigma^0}-m_{\Xi^0}}\right)\frac{1}{2}\left(\sqrt{3}A_{\Lambda p}+A_{\Sigma^+p}\right)$\\
%
$\Xi^0\Sigma^0\eta$&$\left(g^{S}_{\Xi^0\Xi^0\eta}-g^{S}_{\Sigma^0\Sigma^0\eta}\right)\frac{1}{m_{\Sigma^0}-m_{\Xi^0}}\frac{-1}{2\sqrt{2}}\left(A_{\Sigma^+ p}+\sqrt{3}A_{\Lambda p}\right)$\\
%
$\Xi^-\Sigma^-\eta$&$\left(g^{S}_{\Xi^-\Xi^-\eta}-g^{S}_{\Sigma^-\Sigma^-\eta}\right)\frac{1}{m_{\Sigma^-}-m_{\Xi^-}}\frac{1}{2}\left(\sqrt{3}A_{\Lambda p}+A_{\Sigma^+p}\right)$
\end{tabular}}
\end{ruledtabular}
\end{sidewaystable*}

\begin{sidewaystable*}
\caption{\label{taula:resultsWPCVector}Weak baryon-baryon-vector meson parity-conserving couplings, $g^{PC}_{BBV}$.}
\begin{ruledtabular}
\resizebox{\textwidth}{!}{
\begin{tabular}{ccc}
\textbf{Coupling}&\textbf{Analytic value}\\ \hline
%
$pnK^{*+}$&$g^{S}_{\Lambda p K^{*+}}\frac{1}{m_n-m_\Lambda}\frac{-A_{\Lambda p}}{\sqrt{2}}+g^{S}_{\Sigma^0p K^{*+}}\frac{1}{m_n-m_{\Sigma^0}}\frac{-A_{\Sigma^+p}}{\sqrt{2}}$\\
%
$ppK^{*0}$&$g^{S}_{p\Sigma^+\overline{K}^{*0}}\frac{1}{m_p-m_{\Sigma^+}}A_{\Sigma^+p}$\\
%
$nnK^{*0}$&$g^{S}_{n\Lambda\overline{K}^{*0}}\frac{1}{m_n-m_\Lambda}\frac{-A_{\Lambda p}}{\sqrt{2}}+g^{S}_{n\Sigma^0\overline{K}^{*0}}\frac{1}{m_n-m_{\Sigma^0}}\frac{-A_{\Sigma^+p}}{\sqrt{2}}$\\
%
$\Xi^0n\overline{K}^{*0}$&$g^{S}_{\Lambda nK^{*0}}\frac{1}{m_{\Xi^0}-m_{\Lambda}}\frac{1}{\sqrt{2}}\left(A_{\Lambda p}-\sqrt{3}A_{\Sigma^+p}\right) +g^{S}_{\Sigma^0 nK^{*0}}\frac{1}{m_{\Xi^0}-m_{\Sigma^0}}\frac{-1}{2\sqrt{2}}\left(A_{\Sigma^+p}+\sqrt{3}A_{\Lambda p}\right) +g^{S}_{\Xi^0\Lambda K^{*0}}\frac{1}{m_{\Lambda}-m_n}\frac{-A_{\Lambda p}}{\sqrt{2}} +g^{S}_{\Xi^0\Sigma^0 K^{*0}}\frac{1}{m_{\Sigma^0}-m_n}\frac{-A_{\Sigma^+p}}{\sqrt{2}}$\\
%
$\Xi^0pK^{*-}$&$g^{S}_{\Lambda p K^{*+}}\frac{1}{m_{\Xi^0}-m_\Lambda}\frac{1}{\sqrt{2}}\left(A_{\Lambda p}-\sqrt{3}A_{\Sigma^+p}\right)+g^{S}_{\Sigma^0p K^{*+}}\frac{1}{m_{\Xi^0}-m_{\Sigma^0}}\frac{-1}{2\sqrt{2}}\left(A_{\Sigma^+p}+\sqrt{3}A_{\Lambda p}\right)+g^{S}_{\Xi^0\Sigma^+ K^{*+}}\frac{1}{m_p-m_{\Sigma^+}}A_{\Sigma^+p}$\\
%
$\Xi^0\Lambda\rho^0$&$g^{S}_{\Xi^0 \Xi^0\rho^0}\frac{1}{m_\Lambda-m_{\Xi^0}}\frac{1}{\sqrt{2}}\left(A_{\Lambda p}-\sqrt{3}A_{\Sigma^+p}\right)+g^{S}_{\Lambda \Sigma^0\rho^0}\frac{1}{m_{\Xi^0}-m_{\Sigma^0}}\frac{-1}{2\sqrt{2}}\left(A_{\Sigma^+p}+\sqrt{3}A_{\Lambda p}\right)$\\
%
$\Xi^{-}\Lambda\rho^{-}$&$g^{S}_{\Sigma^-\Lambda\rho^+}\frac{1}{m_{\Xi^-}-m_{\Sigma^-}}\frac{1}{2}\left(\sqrt{3}A_{\Lambda p}+A_{\Sigma^+p}\right)+g^{S}_{\Xi^-\Xi^0\rho^+}\frac{1}{m_\Lambda-m_{\Xi^0}}\frac{1}{\sqrt{2}}\left(A_{\Lambda p}-\sqrt{3}A_{\Sigma^+p}\right)$\\
%
$\Xi^0\Lambda\omega$&$\left(g^{S}_{\Xi^0\Xi^0 \omega}-g^{S}_{\Lambda \Lambda\omega}\right)\frac{1}{m_\Lambda-m_{\Xi^0}}\frac{1}{\sqrt{2}}\left(A_{\Lambda p}-\sqrt{3}A_{\Sigma^+p}\right)$\\
%
$\Xi^0\Sigma^0\rho^0$&$g^{S}_{\Xi^0\Xi^0 \rho^0}\frac{1}{m_{\Sigma^0}-m{\Xi^0}}\frac{-1}{2\sqrt{2}}\left(A_{\Sigma^+p}+\sqrt{3}A_{\Lambda p}\right)+g^{S}_{\Sigma^0\Lambda \rho^0}\frac{1}{m_{\Xi^0}-m_\Lambda}\frac{1}{\sqrt{2}}\left(A_{\Lambda p}-\sqrt{3}A_{\Sigma^+p}\right)$\\
%
$\Xi^{-}\Sigma^-\rho^{0}$&$\left(g^{S}_{\Xi^-\Xi^- \rho^0}-g^{S}_{\Sigma^-\Sigma^- \rho^0}\right)\frac{1}{m_{\Sigma^-}-m_{\Xi^-}}\frac{1}{2}\left(\sqrt{3}A_{\Lambda p}+A_{\Sigma^+p}\right)$\\
%
$\Xi^{-}\Sigma^0\rho^{-}$&$\left(g^{S}_{\Sigma^-\Sigma^0\rho^+}\frac{1}{m_{\Xi^-}-m_{\Sigma^-}}+\frac{-1}{\sqrt{2}}g^{S}_{\Xi^-\Xi^0\rho^+}\frac{1}{m_{\Sigma^0}-m_{\Xi^0}}\right)\frac{1}{2}\left(\sqrt{3}A_{\Lambda p}+A_{\Sigma^+p}\right)$\\
%
$\Xi^0\Sigma^0\omega$&$\left(g^{S}_{\Xi^0\Xi^0 \omega}-g^{S}_{\Sigma^0\Sigma^0 \omega}\right)\frac{1}{m_{\Sigma^0}-m_{\Xi^0}}\frac{-1}{2\sqrt{2}}\left(A_{\Sigma^+p}+\sqrt{3}A_{\Lambda p}\right)$\\
%
$\Xi^{-}\Sigma^-\omega$&$\left(g^{S}_{\Xi^-\Xi^- \omega}-g^{S}_{\Sigma^-\Sigma^- \omega}\right)\frac{1}{m_{\Sigma^-}-m_{\Xi^-}}\frac{1}{2}\left(\sqrt{3}A_{\Lambda p}+A_{\Sigma^+p}\right)$
%
%
%
\end{tabular}}
\end{ruledtabular}
\end{sidewaystable*}
\end{document}